\documentclass[12pt,a4paper]{elsarticle}
\usepackage{amsmath}
\usepackage{amsthm}
\usepackage{amssymb}
\usepackage{booktabs}
\usepackage{marvosym}
\usepackage{multirow}
\usepackage{algorithm}
\usepackage{algorithmic}
\usepackage{makecell}
\usepackage{ulem}
\usepackage{oplotsymbl}
\usepackage{rotating}
\usepackage{subfigure}

\newtheorem{definition}{\indent Definition}
\begin{document}
\begin{frontmatter}
    \title{Against Filter Bubbles: Diversified Music Recommendation via Weighted Hypergraph Embedding Learning}
    \author[scu]{Chaoguang Luo}
    \author[swpu]{Liuying Wen}
    \author[scu,polyu]{Yong Qin}
    \author[uic]{Liangwei Yang}
    \author[scu]{Zhineng Hu\corref{cor1}} 
    \author[uic]{Philip S. Yu}

    \address[scu]{Business School, Sichuan University, Chengdu 610064, China}
    \address[swpu]{School of Computer Science and Software Engineering, Southwest Petroleum University, Chengdu 610500, China}
    \address[polyu]{Department of Management and Marketing, The Hong Kong Polytechnic University, Hung Hom, Kowloon, Hong Kong, China}
    \address[uic]{University of Illinois at Chicago, Chicago, USA}

    \cortext[cor1]{Corresponding author. Tel.: +86 135 4038 5354.\\ Email: huzn@scu.edu.cn}

\begin{abstract}
Recommender systems serve a dual purpose for users: sifting out inappropriate or mismatched information while accurately identifying items that align with their preferences.
Numerous recommendation algorithms are designed to provide users with a personalized array of information tailored to their preferences.
Nevertheless, excessive personalization can confine users within a ``filter bubble".
Consequently, achieving the right balance between accuracy and diversity in recommendations is a pressing concern.
To address this challenge, exemplified by music recommendation, we introduce the Diversified Weighted Hypergraph music Recommendation algorithm (DWHRec).
In the DWHRec algorithm, the initial connections between users and listened tracks are represented by a weighted hypergraph.
Simultaneously, associations between artists, albums and tags with tracks are also appended to the hypergraph.
To explore users' latent preferences, a hypergraph-based random walk embedding method is applied to the constructed hypergraph.
In our investigation, accuracy is gauged by the alignment between the user and the track, whereas the array of recommended track types measures diversity.
We rigorously compared DWHRec against seven state-of-the-art recommendation algorithms using two real-world music datasets.
The experimental results validate DWHRec as a solution that adeptly harmonizes accuracy and diversity, delivering a more enriched musical experience.
Beyond music recommendation, DWHRec can be extended to cater to other scenarios with similar data structures.

\end{abstract}
\end{frontmatter}

\section{Introduction}\label{sec: introduction}
Recommender systems (RecSys), as one of the most commonly used methods for technical information filtering, yield tangible benefits for both service providers and users \cite{bradley2001improving, hu2008implicit, gatta2022music}.
In our current digital era, online service providers deploy these systems to deliver personalized suggestions, aiming to heighten user satisfaction by catering to a broad spectrum of needs across diverse user profiles \cite{gatta2022music, yang2023dgrec}.
Through the utilization of recommender systems, service providers can better discern user preferences, boost product visibility, and enhance user engagement \cite{koren2009matrix, rakshit2023popularity}.
For users, recommender systems offer liberation from the overwhelming sea of information and facilitate the discovery of content aligned with their tastes \cite{rendle2012bpr}.
Many recommendation algorithms \cite{he2017neumf, xue2017deep, he2020lightgcn} relentlessly pursue precision in their recommendations, striving to push the boundaries of accuracy to the utmost.

Nevertheless, the excessive pursuit of recommendation accuracy may lead to personalized outcomes that exceed user expectations, thereby exposing them to the pitfalls of filter bubbles \cite{bozdag2014offline, curkovic2019need}.
The idea that filter bubbles \cite{pariser2011filter} shape individuals' thoughts by influencing the information they receive is rooted in communication, informatics, sociology and psychology \cite{curkovic2019need, knobloch2023algorithmic}.
Numerous studies consistently highlight a reality: filter bubbles undeniably exert an impact on users within social networks \cite{dahlgren2021critical, wolfowicz2023examining, michiels2022what}.
An effective approach to bursting the bubble is to implement diversity-aware strategies \cite{resnick2013bursting, schedl2015tailoring, li2021context}.
Studies have documented that recommendation diversity is crucial in many cases, and poor diversity characteristics undermine traditional recommender systems \cite{bradley2001improving, adomavicius2012improving}.


In this paper, we study the diversification problem of music recommendation, which is one of the most prevalent applications
penetrating our daily lives.
It emerges as a critical application of web services, witnessing millions of listening events occurring between a vast user base and tracks in the music repository every moment~\citep{hu2008implicit, shao2009acoustic, rendle2012bpr}.
To obtain an effective diversified music RecSys, we need to consider the characteristics of the music recommendation data.
Firstly, it has rich auxiliary information for each music track, including its Albums, Artists and annotated Tags.
Each auxiliary information reveals one kind of relationship among the music tracks.
For example, music belonging to the same album tends to have similar theme, which reveals the similarity among all music tracks within the same album.
Secondly, music recommendation data is not just one-time interactions.
Unlike user behavior in E-commerce, where there is usually at most one-time interaction between each user-item pair, users can interact multiple times with a single music track.
This characteristic causes each single-time interaction between user-item pairs not to be convincing to speculate the user's preference.
We are unable to assume the user's preference toward a music track if the user only listens to the track one time.
On the contrary, only a one-time interaction usually indicates a disfavor toward the music track because the user never listens to it again.
Thus, we need to discern the user's preference in a more fine-grained manner.

To fully consider the aforementioned characteristics in diversified music recommendation, we introduce the Weighted Hypergraph Convolution for Recommendation (DWHRec).
DWHRec first integrates the relationships between users, tracks, albums, artists and tags by constructing a unified hypergraph.
Specifically, we build $4$ types of hyper-edges within DWHRec to represent the interactions between user-track, tag-track, album-track, and artist-track separately.
For example, each album-track type hyper-edge connects all the music tracks within the same album.
By constructing the hyper-edges within the same hyper-graph, DWHRec is able to fully utilize all the auxiliary information to construct relationships between music tracks.
Besides, to describe user's preference in a more fine-grained manner, we add weight to each hyper-edge based on the edge type.
For example, the weight of each user-item hyperedge is calculated based on the interaction number, more interactions lead to larger edge weight.
In this way, we can discern user's behaviors regarding the interaction number on each item.

After fully representing the data as a hyper-graph, we propose two methods on different steps to diversify the recommendation.
During the embedding step, we first perform random walks on the constructed weighted hypergraph to collect walk sequences, and then utilize a skip-gram model to obtain the user/track embedding.
Compared with graph convolution with fixed aggregation neighborhood, random-walk-based methods can reach the neighborhood at a farther distance, benefiting the diversified representation of center nodes.
Besides, our constructed hypergraph contains rich item relationship information, and random walking on the weighted hypergraph can acquire more diversified walk paths compared with walking on the user-item bipartite graph.
During the ranking step, we also design a novel diversifying function to rank the list.
DWHRec first calculates the relevance score by the dot product of user/track embedding and then re-scores the relevance with a trade-off factor to introduce diversity.
Then, the final ranking list considers both relevance and diversity, easily trading off with the designed factor.
In conjunction with the two designed diversifying methods, DWHRec achieves the best performance on both accuracy and diversity on the real-world music recommendation dataset, yielding its effectiveness on the diversified music recommendation task.
To summarize, the main contributions of this study can be outlined as follows:

\begin{itemize}
    \item Introduction of DWHRec: A novel method that integrates users, tracks, albums, artists, and tags into a unified hypergraph, enhancing the representation of relationships in music recommendation data.
    \item We propose random walks on the weighted hypergraph to generate more diversified walk paths based on different sources of information. Combined with the skip-gram model, DWHRec can obtain more diversified user/track embeddings.
    \item In the ranking stage, we introduce a novel diversifying function that calculates relevance scores adjusted by a trade-off factor to balance relevance and diversity in the final recommendation list.
    \item Extensive experiments are conducted on real-world music datasets to demonstrate the effectiveness of DWHRec and the influences of different modules. Comparative experiments show that DWHRec is capable of striking an appropriate balance between recommendation accuracy and diversity.
\end{itemize}

The remainder of this study is organized as follows.
Firstly, we present a brief review of existing related works in the fields of music recommendation systems, diversified recommendation and hypergraph-based recommendation (Section \ref{sec: related-work}).
Secondly, we construct a diversified music recommendation algorithm based on weighted hypergraph embedding (Section \ref{sec: weighted-hypergraph-based-diversity-music-recommendation}).
Thirdly, in the experimental session, we compare the proposed DWHRec with several state-of-the-art recommendation algorithms, and detailed experimental results are reported (Sections \ref{sec: experimental-setup}--\ref{sec: experimental-results}).
Finally, in the concluding part of this study, we summarize the findings and envision possible extensions (Section \ref{sec: conclusions}).

\section{Related Work}\label{sec: related-work}
As stated afore, we leverage music recommendations to investigate filter bubbles and employ a diversified recommendation approach to alleviate the problem.
Therefore, it is essential to introduce the related concepts and the current state of research.
Our study aligns with three main research directions: music recommendation, diversified recommendation and hypergraph-based recommendation.

\subsection{Music Recommendation}
In this study, our focus is on music recommender systems.
With the proliferation of digital music, the evolution of music recommendation proves beneficial for users in picking desirable music pieces from an extensive repository \cite{shan2009emotion}.
The various music recommendation methods developed thus far can be broadly classified into two fundamental categories: content-based and collaborative-filtering approaches.

Music is an artistic presentation of sound and is characterized by numerous acoustic features \citep{shao2009acoustic,tan2011rich,cheng2016acoustic,niyazov2021content,sakurai2022deep}.
Huang \cite{huang2004audio} initially extracted audio signatures as music features from audio data, rated new music by utilizing a vector quantization method, and ultimately generated music recommendations.
In contrast to recommendations that solely focus on the music content itself, Bu \cite{bu2010music} incorporated features extracted from the Mel-frequency cepstral coefficients as a bridge to compare similarities between tracks, embedding them into a unified hypergraph architecture.

Music typically conveys some form of emotion in addition to its acoustic characteristics.
Shan \cite{shan2009emotion} explored the discovery of affinity in film music and proposed a generalized framework for implementing emotion-based music recommendations.
To tailor recommendations to better suit the user's current context, Hariri \cite{hariri2012using} mined popular tags for tracks from social media websites, employed a topic modeling approach to learn latent topics representing various contexts from these tags, and then transformed each track into a set of latent topics capturing the general characteristics of that track.

Another strategy, which is the focus of our attention in this study, relies solely on the past behavior of the user, excluding the content of the music itself, often difficult to access.
The direct approach refers to detecting associations between users, tracks, albums and artists through available information, uncovering the latent structure between them, and deducing the tracks that users may be attracted to, providing recommendations \cite{schedl2015tailoring}.
Mao \cite{mao2016quality} corrected user preference relations found from users' ratings with a quality model and proposed a regularization framework to calculate the recommendation probability of tracks.
Knowledge graphs, as tools to reason on data to extract new and implicit information, were naturally applied to music recommendations that mine associations between users and tracks \cite{oramas2016sound}.
La Gatta et al. \cite{gatta2022music} suggested that hypergraph data models might be more capable of seamlessly representing all possible and complex interactions between users and tracks with related characteristics.
Furthermore, music play has a natural characteristic, i.e., sequence.
Specifically, Cheng \cite{cheng2017exploiting} exploited this property to seek the relevance of tracks and attempted to leverage the information encoded in music play sequence into the matrix factorization methods \citep{koren2009matrix, loepp2019interactive} to improve the recommendation performance.

In summary, content-based music recommendation aims to establish similarities between tracks, with less consideration for personalization, while collaborative filtering-based recommendation pays more attention to detecting potential associations between users and tracks, being more inclined to match the users' historical preferences.

\subsection{Diversified Recommendation}
Over the years, numerous recommendation algorithms have been developed to construct recommender systems that process massive amounts of data, aiming to identify potential user preferences and provide optimal recommendations \citep{resnick1994grouplens, balabanovic1997fab, deshpande2004item, zhao2016improved}.
Ongoing studies on recommender systems often prioritize accuracy as the sole objective \citep{wu2023survey, wang2022attention, zhang2023dynamicgraph}.
However, most personalized recommender systems compromise the accuracy of recommended items in their pursuit of increased diversity in recommendations \citep{fleder2007recommender, heitz2022democracy}.
Simultaneously, it has been pointed out that accuracy may not be the exclusive goal of a recommender system \citep{mccrae1992introduction, lu2018diversity, xie2022improving}.
Recommender systems would benefit from giving more consideration to other crucial aspects, such as diversity and novelty.
Research has shown that, frequently, the diversity of recommendations is significant, and traditional recommender systems tend to exhibit poor diversity characteristics \citep{bradley2001improving, adomavicius2012improving, huang2019item, heitz2022democracy}.

Diversified recommendations can enhance the diversity and serendipity of a recommendation list, providing surprises and satisfaction to users, as these items can remain highly relevant to users \citep{liu2020long, robinson2020user}.
Furthermore, diversified recommendations can assist in exploring long-tail items, which is another challenging area and hot spot within recommender systems research \citep{liu2020long}.

Many current diversity-oriented recommender systems adopt a fixed strategy to adjust the diversity degree for all users.
Typically, they define a score function that balances diversity and accuracy with a hyper-parameter.
Subsequently, the generated recommendation list is re-ranked based on the calculated scores \citep{lu2018diversity, robinson2020user}.
For instance, Bradley \cite{bradley2001improving} focused on a variation of a quality metric achieved through a linear combination of similarity and diversity, where the relative weight of the similarity and diversity factors can be altered by adjusting a hyper-parameter.
Simultaneously, a non-linear alternative form of the quality metric, computed as the simple harmonic mean of similarity and diversity, has been proposed.

Beyond the pre-defined score function mentioned above, multi-objective optimization technology offers a viable alternative for balancing accuracy and diversity \citep{ma2023research}.
Unlike single-objective optimization, which can optimize only one objective function, multi-objective optimization can concurrently optimize several objective functions to obtain the Pareto optimal solution set.
Two indicators, intra-user diversity and mean absolute error were selected to evaluate recommended diversity and accuracy, respectively \citep{ma2023research}.

Taken together, this simple way of defining diversity does not apply to more complex scenarios, such as music recommendation, where relationships between multiple entities need consideration.

\subsection{Hypergraph-based Recommendation}
Recently, numerous studies have delved into hypergraphs \citep{yang2019revisiting, pavone2022online, yang2023group}.
For instance, Bu \cite{bu2010music} introduced a unified hypergraph framework for music recommendation, incorporating diverse social media information and acoustic-based content into the algorithm.
Theodoridis \cite{theodoridis2013group} extended Bu's framework \cite{bu2010music} to include group sparsity constraints, enabling the exploitation of the group structure within the data.
Treating music recommendation as a hypergraph-based ranking problem, Tan \cite{tan2011rich} integrated rich social media information to identify music tracks tailored to individual user preferences.
A novel music recommendation framework that leverages a hypergraph data model and hypergraph embedding techniques was delivered by La Gatta \cite{gatta2022music}.
By reexamining user mobility and social relationships, a hypergraph embedding method is specifically designed for a large-scale location-based social network dataset, facilitating automatic feature learning \cite{yang2019revisiting}.
Introducing a generic user-item-attribute-context data model summarizing various information resources and higher-order relationships for constructing a multipartite hypergraph fulfilling multi-objective recommendation needs, a solution was proposed, utilizing hypergraph ordering \citep{mao2019multiobjective}.

While these approaches have garnered considerable success in resource recommendation applications, they have yet to establish a centralized hypergraph.
Our goal is to construct a hypergraph with items at the center, surrounded by other resources, aiming to enhance the strength of connections between resources.

\section{Weighted Hypergraph-based Diversified music recommendation}
\label{sec: weighted-hypergraph-based-diversity-music-recommendation}
The next key point for discussion involves around the method of seeking and establishing connections among users, tracks, albums, artists and tags.
Utilizing the hypergraph, we present a music recommendation algorithm based on hypergraph embedding, meticulously designed to strike a balance the weights of accuracy and diversity.
This section provides a detailed illustration of our approach.

\subsection{Preliminaries}
We summarize the notation in Table \ref{tab: notation}.
The graph comprises two essential components: vertices and edges.
A hypergraph, denoted as $\mathcal{G} = (\mathcal{V}, \mathcal{E})$, consists of a vertex set $\mathcal{V}$ and a hyperedge set $\mathcal{E}$.
Each hyperedge $e \in \mathcal{E}$ encompasses an arbitrary number of vertices $v \in \mathcal{V}$.
For a weighted hypergraph, represented as $\mathcal{G} = (\mathcal{V}, \mathcal{E}, \rm{w})$, the vertices $\mathcal{V}$ and hyperedges $\mathcal{E}$ are accompanied by a weighting function $f_{\rm{w}}(e)$: $\mathcal{E} \rightarrow \mathbb{R}$,  indicating the strength of connections.
A higher weight signifies closer proximity between the connected vertices.
Additionally, each hyperedge $e \in \mathcal{E}$ is linked to a non-negative number $\rm{w}(e)$, referred to as the weight of the hyperedge $e$.

A hyperedge $e$ is said to contain a vertex $v$ when $v \in e$.
The degree of a hyperedge $e$, denoted by $\delta(e)$, is defined as the cardinality of $e$, i.e., $\delta(e)=\left| e \right|$.
If every hyperedge has a degree of $2$, the hypergraph reduces to a normal graph.
A hyperpath between vertices $v_1$ and $v_k$ exists when there is an alternative sequence of distinct vertices and hyperedges $v_1$, $e_1$, $v_2$, $e_2$, $\dots$, $e_{k-1}$, $v_k$ such that $\left\{v_i, v_{i+1}\right\}  \subseteq e_i$ for $1 \leq i \leq k-1$.
A hypergraph is connected if there is a path for every pair of vertices.

Finally, we obtain the vertex-hyperedge incidence matrix, $\mathbf{H}$.
The hypergraph $\mathcal{G}$ can be succinctly represented by $\mathbf{H}$ of size $\left|\mathcal{V} \right| \times \left| \mathcal{E} \right|$ matrix.
Let $h(v, e)=1$ indicate that a vertex $v$ is part of a hyperedge $e$, and $h(v,e)=0$ otherwise.
The incidence matrix, $\mathbf{H}$, is defined by its elements:
\begin{equation}\label{equ: incidence-matrix}
  H_{ij}=h(v_i,e_j)=
  \begin{cases}
    1, & \mbox{if } v_i\in e_j; \\
    0, & \mbox{otherwise}.
  \end{cases}
\end{equation}
Logically, we define the degree of a vertex as
\begin{equation}\label{equ: vertex-degree}
  d(v)=\sum_{e}{\rm{w}}(e)h(v,e),
\end{equation}
and the degree of a hyperedge as the number of connected vertices, denoted as,
\begin{equation}\label{equ: hyperedge-degree}
  \delta(e)=\left| e \right|=\sum_{v} h(v,e).
\end{equation}

\begin{table}
  \centering
  \caption{Hypergraph notations}\label{tab: notation}
  \setlength{\tabcolsep}{10mm}
  \begin{tabular}{ll}
    \toprule
    \textbf{Symbol} & \textbf{Description} \\
    \midrule
    $\mathcal{G} = (\mathcal{V}, \mathcal{E}, \rm{w})$ & Hypergraph. \\
    $\mathcal{V}$ & The set of vertices. \\
    $\mathcal{E}$ & The set of hyperedges. \\
    $\rm{w}$ & The weight on hyperedges and vertices.\\
    $v \in \mathcal{V}$ & A certain vertex within the vertex set. \\
    $e \in \mathcal{E}$ & A certain hyperedge within the hyperedge set. \\
    $\delta(\cdot)$ & The degree of a vertex or a hyperedge.  \\
    $\mathbf{H}$ & The incidence matrix. \\
    $h$ & The elements in the incidence matrix. \\
    $U \subseteq \mathcal{V}$ & The set of users. \\
    $Tr \subseteq \mathcal{V}$ & The set of tracks. \\
    $Ar \subseteq \mathcal{V}$ & The set of artists. \\
    $Al \subseteq \mathcal{V}$ & The set of albums. \\
    $Ta \subseteq \mathcal{V}$ & The set of tags. \\
    $u \in U$ & A certain user within the user set. \\
    $tr \in Tr$ & A certain track within the track set. \\
    $al \in Al$ & A certain album within the album set. \\
    $ar \in Ar$ & A certain artist within the artist set. \\
    $ta \in Ta$ & A certain ta within the tag set. \\
    $Le$ & The listening history.\\
    $c$  & The quantity of elements. \\
    \bottomrule
  \end{tabular}
\end{table}

\subsection{Hypergraph Composition}\label{subsec: hypergraph-composition}
The hypergraph structure is a fundamental component of our proposed algorithm.
Consequently, it is essential to furnish a detailed configuration explanation for the elements within the hypergraph.

We consider five types of objects and four types of relationships.
Specifically, the objects comprise users $U$, three resource types (i.e., tracks $Tr$, albums $Al$ and artists $Ar$) and tags $Ta$ that users attach to them.
The relationships within the constructed hypergraph are partitioned into actions on resources and inclusion relationships among resources.
Actions relations on resources engage two types of interactions: users listening to tracks ($R_1$) and users tagging tracks ($R_2$).
It is worth noting that the relationship between users and the tagging of tracks represents the collective annotation of a particular track by all users, rather than a specific user.
Inclusion relationships among resources are defined by connections between tracks and releases and between tracks and artists, signified by $R_3$ and $R_4$, respectively.

The hypergraph $\mathcal{G}$ is defined as a collection of vertices $\mathcal{V}$ and hyperedges $\mathcal{E}$.
The vertex set, $\mathcal{V}$, comprises distinct entities, specifically:

(1) \emph{Users} ($U$): the set of users;

(2) \emph{Albums} ($Al$): the set of albums;

(3) \emph{Artists} ($Ar$): the set of artists;

(4) \emph{Tracks} ($Tr$): the set of tracks;

(5) \emph{Tags} ($Ta$): the set of tags that users attach to tracks.

So, the set of vertices $\mathcal{V}$, is defined as the union of $U$, $Al$, $Ar$, $Tr$ and $Ta$, i.e., $\mathcal{V}=U \cup Al \cup Ar \cup Tr \cup Ta$.
Hyperedges are introduced to represent the four relationships among the aforementioned objects.
In the unified hypergraph, there exist four types of hyperedges, each corresponding to a specific relationship type.
To correspond with the relation set $R_i$, we define the set of hyperedges as $e^{(i)}$, ($i \in [1,4], i \in \mathcal{R}^{+}$).
For convenience and better comprehension, we provide an illustrative representation of these relationships in Table \ref{tab: relations-in-our-unified-hypergraph}.

\begin{table}[htp]
  \centering
  \setlength{\tabcolsep}{5mm}
  \caption{Relations in our unified hypergraph}\label{tab: relations-in-our-unified-hypergraph}
  \begin{tabular}{lcc}
    \hline
    \toprule
    \multirow{2}*{Name} & \multicolumn{2}{c}{Notations} \\
    \cline{2-3}
      & Relations & Hyperedge Types \\
    \midrule
    Listening tracks             & $R_1$    & $e^{(1)}$  \\
    Tagging tracks               & $R_2$    & $e^{(2)}$  \\
    Tracks belong to an album    & $R_3$    & $e^{(3)}$  \\
    Tracks belong to an artist   & $R_4$    & $e^{(4)}$  \\
    \bottomrule
    \hline
  \end{tabular}
\end{table}

The construction of the four types of relations and hyperedges is listed as follows.
\begin{definition}[Hyperedge $e^{(1)}$]\label{def: hyperedge-user-track}
The first type of hyperedge, denoted as $e^{(1)} \in \mathcal{E}$, represents the tracks that the user has listened to in the past.
It consists of two parts: the user vertex and track vertices.
The play count, a significant metric for listening events, is denoted as $c(u, tr_j)$, where $u \in U$ is the user and $tr_j \in Tr$ is the track.
In $e^{(1)}$, the weight of the user vertex $v_{u} \in \mathcal{V}$ is set to $1$, while the weight of other track vertex $v_{tr_j}$ is determined as follows:
\begin{equation}\label{equ: weight-of-listening}
  {\rm{w}}(v_u, v_{tr_j}) = \frac{c(u, tr_j)}{\sum_{tr_i} c(u, tr_i)},
\end{equation}
where $tr_i$ represents the track belonging to the same hyperedge as user $u$.
\end{definition}
\begin{definition}[Hyperedge $e^{(2)}$]
\label{def: hyperedge-tag-track}
The second type of hyperedge, denoted as $e^{(2)} \in \mathcal{E}$, represents the tags that the user has annotated to a track.
It comprises two components: the track vertex and the attached tag vertices.
In $e^{(2)}$, the number of times the tag $ta_p \in Ta$ attached to the track $tr \in Tr$ is denoted as $c(tr, ta_p)$.
The weight of the track vertex $v_{tr}$ is set to $1$, and the weight of other tag vertex $v_{ta_p}$ is set to:
\begin{equation}\label{equ: hypergraph-weight-of-tag}
  {\rm{w}}(v_{tr}, v_{ta_p}) = \frac{c(tr, ta_p)}{\sum_{ta_i} c(tr, ta_i)},
\end{equation}
where $ta_i$ represents the tag belonging to the same hyperedge as track $tr$.
\end{definition}
\begin{definition}[Hyperedge $e^{(3)}$]
\label{def: hyperedge-album-track}
Certain connections exist between different tracks within the same album, prompting the natural construction of a hyperedge for tracks belonging to the same album.
All the tracks mentioned here that share an album constitute the albums in the dataset.
The third type of hyperedge, denoted as $e^{(3)} \in \mathcal{E}$, represents tracks belonging to the same album.
It contains two components: the album vertex and the tracks vertices it includes.
In $e^{(3)}$, the notation $c(al, tr_j)$ denotes the number of times the track $tr_j \in Tr$ is played in the album $al \in Al$.
The weight of the album vertex $v_{al}$ is set to $1$, and the weight of other track vertex $v_{tr_j}$ is set to:
\begin{equation}\label{equ: hypergraph-weight-of-album}
  {\rm{w}}(v_{al}, v_{tr_j}) = \frac{c(al, tr_j)}{\sum_{tr_i} c(al, tr_i)},
\end{equation}
where $tr_i$ refers to the track that belongs to the same hyperedge as album $al$.
\end{definition}
\begin{definition}[Hyperedge $e^{(4)}$]
\label{def: hyperedge-artist-track}
Similarly, hyperedges can be also created for tracks and artists.
Multiple tracks are composed or performed by the same artist.
The hyperedge $e^{(4)}$ captures the relationship between the artist and several tracks, designed to prevent the omission of important tracks and serving as a complement to $e^{(3)}$.
The vertices of $e^{(4)}$ encompass all tracks belonging to the artist and the artist itself.
In $e^{(4)}$, the frequency with which the track $tr_j \in Tr$ by the artist $ar \in Ar$ has been played is denoted as $c(ar, tr_j)$.
The weight of the artist vertex $v_{ar}$ is set to $1$, and the weight of other track vertex $v_{tr_j}$ is regarded as:
\begin{equation}\label{equ: hypergraph-weight-of-artist}
  {\rm{w}}(v_{ar}, v_{tr_j}) = \frac{c(ar, tr_j)}{\sum_{tr_i} c(ar, tr_i)},
\end{equation}
where $tr_i$ is the track belonging to the same hyperedge as artist $ar$.
\end{definition}

\subsection{Recommendation via Hypergraph}
\label{subsec: music-recommendation-via-hypergraph}
\begin{figure}[htp]
  \centering
  \includegraphics[width=\textwidth]{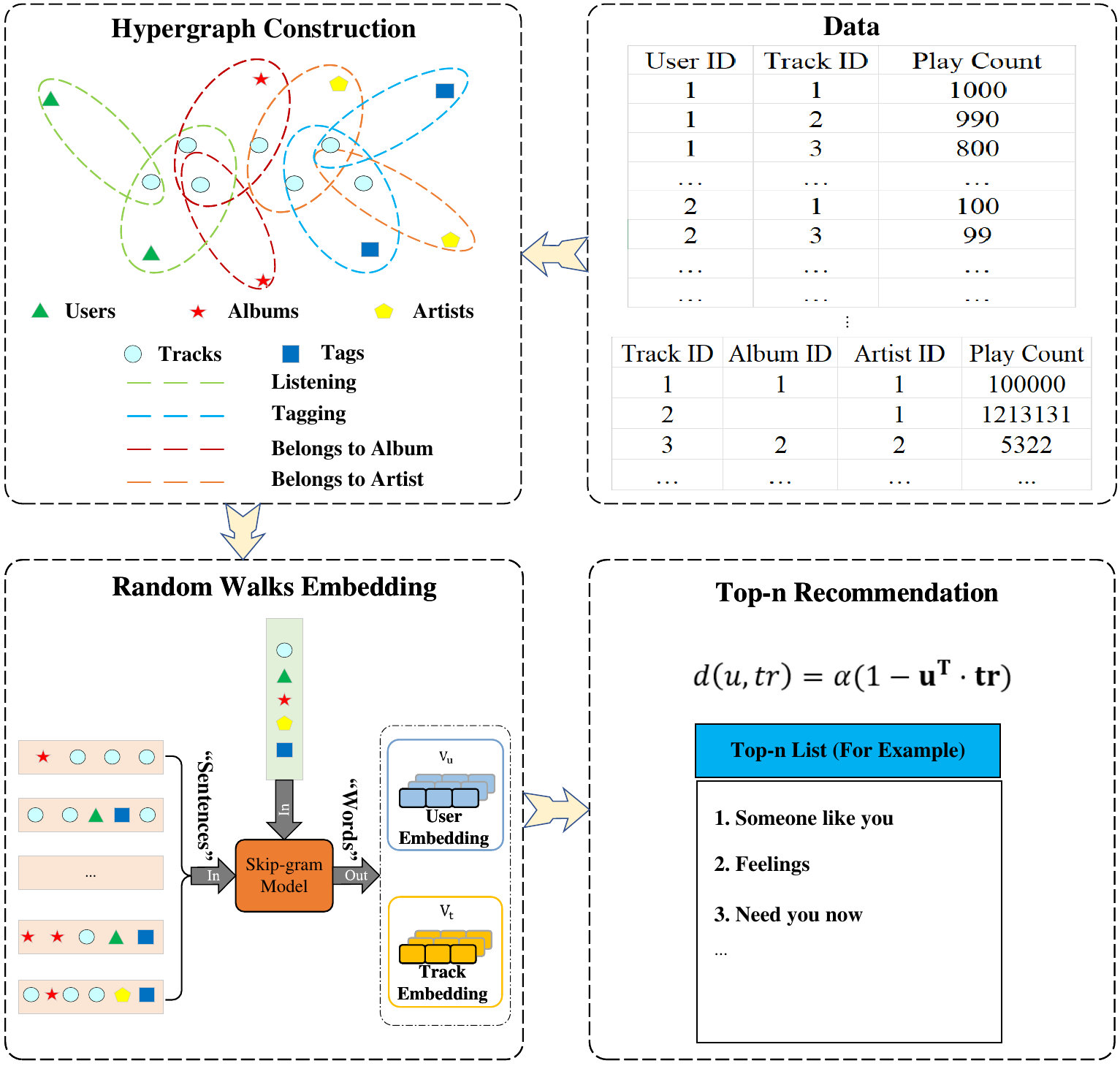}
  \caption{The framework of the DWHRec algorithm. We first construct a hypergraph using the user's historical interactions and external knowledge, such as tags, albums and artists. Subsequently, a random-walks-based embedding method is employed to learn dense vector representations for users and items, facilitating top-$n$ recommendations.}
  \label{fig: framework}
\end{figure}

Figure \ref{fig: framework} vividly and succinctly depicts the entire workflow of the DWHRec algorithm, as detailed in Algorithm \ref{algorithm: framework}, which outlines the core framework of the system.
DWHRec takes two inputs: data and hyper-parameters.
The data includes the user's listening history $Le$, tracks $Tr$, tags $Ta$ associated with the tracks, albums $Al$ and artists $Ar$.
DWHRec receives and processes these inputs through a series of steps to generate lists of recommendations $L$ for all users.
The hyper-parameters consist of the iteration counts $r$ and the number of steps $k$ in the random walks stage, as well as the vector dimension $s$ and window size $w$ in the embedding stage and the recommended list length $n$.

DWHRec can be segmented into four stages (Algorithm \ref{algorithm: framework}).
Firstly, a weighted hypergraph $\mathcal{G}$ is constructed based on information from $Le$, $Tr$, $Ta$, $Al$ and $Ar$ (line $1$).
A more detailed description of this process is presented in Algorithm \ref{algorithm: hypergraph-construction}.
Secondly, utilizing the constructed hypergraph, the random walks method is designed to detect potential associations between users $U$ and candidate tracks $Tr$ (line $2$).
A more detailed explanation of this process is provided in Algorithm \ref{algorithm: random-walks-generation}.
Thirdly, considering the paths left by the vertex walks as sentences and the vertices as words, the skip-gram word embedding model is applied to vectorize the users and tracks vertices, obtaining their vector representations (line $3$).
Finally, the designed scoring function takes the vectors of users and tracks as input, calculating the scores for candidate tracks.
These candidate tracks are graded, and the recommendation list is generated by ranking them according to the highest rating (lines $4$).
Through these steps, DWHRec achieves the goal of providing users with as diversified recommendations as possible.

\begin{algorithm}[htp]
	\caption{DWHRec Framework}
    \label{algorithm: framework}
	\raggedright
	\textbf{Input: }{Data about listening history behaviors $Le$, tracks $Tr$, tags $Ta$, albums $Al$ and artists $Ar$, hyper-parameters include the number of iterations $r$ for random walks, the number of steps $k$ taken during each iteration of random walks, the size $s$ of the vertex vectors generated by word embedding model, the window size $w$ and the length of the recommendation list $n$}.  \\
	\textbf{Output: }{Top-$n$ recommendation list $L$ for all users}. \\
	\textbf{Method: }{DWHRec}.
	
	\begin{algorithmic}[1]
        \STATE{Construct a weighted hypergraph $\mathcal{G}$ by utilizing the information from $Le$, $Tr$, $Ta$, $Al$ and $Ar$ (Algorithm \ref{algorithm: hypergraph-construction})};
        \STATE{Explore possible connections among $U$ and $Tr$ by applying random walks technique on hypergraph $\mathcal{G}$ (Algorithm \ref{algorithm: random-walks-generation})};
        \STATE{Utilize the skip-gram word embedding model to vectorize the vertices in $U$ and $Tr$};
        \STATE{Generate a recommendation list $L$ with a length of $n$ for $U$};
		\RETURN{$L$};
	\end{algorithmic}
\end{algorithm}

\subsubsection{Hypergraph Construction}
\label{subsubsec: hypergraph-construction}
Our objective is to provide users with specific track recommendations in the form of a personalized list.
When making recommendations, the goal is to cater to users' individual preferences by suggesting tracks that are as familiar or similar as possible to their historical listening behaviors.
To assist users in discovering tracks they are likely to enjoy, we model the information using a hypergraph data structure.
This structure captures relationships between various entities, interconnecting all the information embedded in the data source.

As mentioned earlier, a fundamental step in DWHRec is the construction of the hypergraph,
and Algorithm \ref{algorithm: hypergraph-construction} gives the pseudo-code for this process.

\begin{algorithm}[htp]
	\caption{Hypergraph Construction}
    \label{algorithm: hypergraph-construction}
	\raggedright
	\textbf{Input: }{Information about listening history behaviors $Le$, tracks $Tr$, tags $Ta$, albums $Al$ and artists $Ar$};  \\
	\textbf{Output: }{$\mathcal{G}=(\mathcal{V}, \mathcal{E}, \rm{w})$}; \\
	\textbf{Method: }{HypergraphConstruction}.
	
	\begin{algorithmic}[1]
        \STATE{$\mathcal{V} \leftarrow \emptyset$, $\mathcal{E} \leftarrow \emptyset$, $\rm{w} = 0$};
        \STATE{Construct hyperedge $e^{(1)}$ (Definition \ref{def: hyperedge-user-track}), assign weights to vertices (Equation \eqref{equ: weight-of-listening})};
        \STATE{Construct hyperedge $e^{(2)}$ (Definition \ref{def: hyperedge-tag-track}), assign weights to vertices (Equation \eqref{equ: hypergraph-weight-of-tag})};
        \STATE{Construct hyperedge $e^{(3)}$ (Definition \ref{def: hyperedge-album-track}), assign weights to vertices (Equation \eqref{equ: hypergraph-weight-of-album})};
        \STATE{Construct hyperedge $e^{(4)}$ (Definition \ref{def: hyperedge-artist-track}), assign weights to vertices (Equation \eqref{equ: hypergraph-weight-of-artist})};
        \STATE{$\mathcal{V} \leftarrow \mathcal{V} \cup U \cup Ar \cup Al \cup Tr \cup Ta$};
        \STATE{$\mathcal{E} \leftarrow \mathcal{E} \cup e^{(1)} \cup e^{(2)} \cup e^{(3)} \cup e^{(4)}$};
        \STATE{$\mathcal{G} \leftarrow (\mathcal{V}, \mathcal{E}, \rm{w})$};
        \RETURN{$\mathcal{G}$};
	\end{algorithmic}
\end{algorithm}

The hypergraph $\mathcal{G}$ is defined by a triple that comprises the set of vertices $\mathcal{V}$ and hyperedges $\mathcal{E}$, along with their corresponding weights $\rm{w}$.
Naively, hypergraph construction process can be approximately divided into three steps, constructing the vertices set $\mathcal{V}$, constructing the hyperedges set $\mathcal{E}$ and assigning the hyperedges weights $\rm{w}$.
In practice, hyperedges and weights are often combined, with an initial weight being attached to a hyperedge during its construction.

An empty hypergraph $\mathcal{G}$ is initialized with an empty set of hyperedges $\mathcal{E}$ and vertices set $\mathcal{V}$, all assigned a weight of $0$ (line $1$).
Further, the algorithm proceeds to construct the hyperedges $e^{(1)}$, $e^{(2)}$, $e^{(3)}$ and $e^{(4)}$ to represent user-track, tag-track, album-track and artist-track associations, respectively, assigning each corresponding hyperedge an initial weight (lines $2\text{-}5$).
The vertices set $\mathcal{V}$ is formed by concatenating the sets of users $U$, artists $Ar$, albums $Al$, tracks $Tr$ and tags $Ta$ (line $6$).
Additionally, the hyperedges set $\mathcal{E}$ is the union of the hyperedges $e^{(i)}$, where $i$ ranges from $1$ to $4$ (line $7$).
Finally, a hypergraph $\mathcal{G}$ is created (line $8$).

In the constructed hypergraph, tracks assume a central position.
Through these tracks, connections are established among users, tags, albums and artists, facilitating the integration of information.
This structure enables the exploration of users' latent preferences by intertwining information from multiple aspects.

\subsubsection{Random Walks on Hypergraph}
\label{subsubsec: random-walks-on-hypergraph}
After the construction of the hypergraph, the desire is to utilize it for recommendations.
However, the information stored in the hypergraph cannot be used directly, necessitating further processing.
Random walks are conducted on the hypergraph to facilitate the establishment of connections between various entities.

\begin{algorithm}[htp]
	\caption{Random Walks Generation}
    \label{algorithm: random-walks-generation}
	\raggedright
	\textbf{Input: }{Hypergraph $\mathcal{G}$, the number of iterations $r$ for random walks, the number of steps $k$ taken during each iteration of random walks}.  \\
	\textbf{Output: }{Walks list $walks\_list$}. \\
	\textbf{Method: }{RandomWalk}.
	
	\begin{algorithmic}[1]
        \STATE{$walks\_list = \emptyset$};
        \FOR{$v \in \mathcal{V}$}
            \STATE{$walk = \emptyset$};
    		\STATE{$v_{curr} \leftarrow v$};
            \STATE{$e_{curr} \leftarrow e \in \mathcal{E}: v_{curr} \in \mathcal{V}_e $};
            \FOR{$i = 1 \to r$}
                \FOR{$j = 0 \to k$}
                    \STATE{$walk \leftarrow walk + v_{curr}$};
                    \STATE{$ e_{curr} \leftarrow e \in \mathcal{E}: v_{curr} \in \mathcal{V}_e $};
                    \STATE{$ v_{curr} \leftarrow v \in \mathcal{V}_{e_{curr}}, v \neq v_{curr} $};
                \ENDFOR
                \STATE{$walks\_list \leftarrow walk\_list + walk$};
    		\ENDFOR
    	\ENDFOR
    	\RETURN {$walks\_list$};
	\end{algorithmic}
\end{algorithm}

The pseudocode of the random walks process is approximately rendered in Algorithm \ref{algorithm: random-walks-generation}.
The algorithm accepts a hypergraph $\mathcal{G}=(\mathcal{V}, \mathcal{E}, \rm{w})$ as input, along with two hyper-parameters: $r$, representing the number of iterations, and $k$, indicating the number of random walk steps.
It outputs a list containing the walking results for all vertices.
The variable $walks\_list$ is a list that holds the results of all vertices walks, initialized as empty (line $1$).
For each vertex $v \in \mathcal{V}$, a random walk operation is performed, and the order of the walks is recorded (lines $2\text{-}14$).

Before commencing the actual walking process, some preparation are necessary.
Specifically, a variable $walk$ is declared and initialized to empty, tracking the order in which $v$ travels (line $3$).
Let the currently visiting vertex $v_{curr}$ be $v$ (line $4$).
A hyperedge $e \in \mathcal{E}$ is randomly picked from the hyperedges containing the vertex $v_{curr}$ and marked as the currently accessed hyperedge $e_{curr}$ (line $5$).

After the preparation is completed, the actual walking can commence.
The vertex $v_{curr}$ becomes the first vertex visited, and it is appended to $v$'s visited path $walk$ (line $8$).
For the remaining $k\text{-}1$ steps, the following manipulations are performed (lines $9\text{-}10$).
The transition probability $p_{e_{curr}}$ for a hyperedge $e_{curr}$ to jump to another hyperedge is evaluated first.
Based on this probability, the current hyperedge could either remain unchanged or switch to another hyperedge where the current vertex is (line $9$).
This mechanism enables the algorithm to explore more hyperedges deeply, avoiding getting stuck in a loop inside the hyperedges.
After determining the hyperedge for the next walk, the next vertex for the walk needs to be further identified.
The process involves selecting a new vertex $v$ to join the random walk for the next walk based on the probabilities associated with the vertices in the hyperedge $e_{curr}$ (line $10$).
Through these procedures, a single random walk path for vertex $v$ is recorded after $k$ jumps (lines $7\text{-}11$).
The process of $k$ jumps repeated $r$ times iteratively forms the complete walking trajectory for vertex $v$ (lines $6\text{-}13$).
Apparently, the walking paths of all vertices constitute the $walks\_list$ (lines $2\text{-}14$).

The random walks process on the hypergraph generates walking paths for vertices.
Adjacent vertices in the hypergraph are likely to be adjacent in the paths, and non-adjacent vertices are also included in the walking paths, even though they may be separated by larger distances.

\subsubsection{Embedding and Recommendation}
\label{subsubsec: embedding-and-recommendation}
The random walks process transforms the hypergraph structure into walking paths for vertices.
However, further exploration is still required, as the latent preference information of users for tracks remains concealed within these paths.

To be aware of users' track preferences, we employ the skip-gram model \cite{perozzi2014deepwalk} to learn the efficient estimation of word representations in vector space.
Word2Vec \cite{mikolov2013efficient, rong2014word2vec, mikolov2013exploiting} is a class of neural network models that, within the context of a given unlabeled corpus, generates vectors for words in the corpus to characterize their semantic information.
A widely used model within Word2Vec is skip-gram, extensively employed in natural language processing (NLP) as an unsupervised model for learning semantic knowledge from massive text corpora.
In our case, the vertices of the hypergraph are considered as words, while the random walks are treated as sentences.

After the application of the skip-gram model, the users and tracks in the hypergraph are demonstrated as vectors.
Further, the user's enjoyment of a track can be proxied by calculating the dot product between these two vectors.
The well-known dot product metric can be delegated to portray user $u$'s favoring of track $tr$, denoted as:
\begin{equation}\label{equ: cosine-similarity}
  rel(u, tr) = \mathbf{u}^{\text{T}} \cdot \mathbf{tr}.
\end{equation}
The users' preferences for the tracks are then converted into products between the user vertex vector and the vectors associated with the tracks.

Diversity characteristics serve as a measure of how diversified the observed range of users' music is---the diversity of music listened to, intuitively related to openness \cite{farrahi2014impact}.
In general, diversity is defined as the degree of differentiation between items in the recommended list, and is formally expressed as:
\begin{equation}\label{equ: weighting-importance-degree}
  d(u, tr) = \alpha \left(1-rel(u, tr)\right),
\end{equation}
where $\alpha$ is a hyper-parameter that controls the acceptable level of diversity, acting as a trade-off between accuracy and diversity.

In summary, DWHRec algorithm utilizes a hypergraph structure to model the associations between users and tracks.
Subsequently, through multiple iterations of random walks on the hypergraph, it generates traversal paths for both users and tracks.
Next, the skip-gram word embedding model is employed to analyze these traversal paths, representing users and tracks as vectors.
Finally, by computing the diversity degree, the algorithm measures the extent of user preferences for tracks, thereby generating a diversified recommendation list.

\section{Experimental Setup}\label{sec: experimental-setup}
The experimental work to validate the effectiveness of the algorithm is necessary, and for this purpose, extensive, intricate and prolonged preparations have been carried out.
To ensure the relative fairness and transparency of the experiments, we will provide a relatively detailed description of the experimental setup.

\subsection{Dataset Collection}
Last.fm\footnote{http://www.last.fm/} is a well-known data gathering site widely used in the field of music recommendation research \cite{bu2010music, theodoridis2013group, oramas2016sound, robinson2020user}.
It builds a detailed profile of each user's musical tastes by recording the tracks they have historically interacted with.
Last.fm collects these tracks from Internet radio stations or the user's computer, for example, by transferring (``scrubbing") them to Last.fm's database via a music player (e.g. Spotify\footnote{https://open.spotify.com/}) or a plug-in installed in the user's music player.

Users were discovered by crawling the Last.fm social graph using the ``user.getFriends" endpoint and by crawling the listening users of a certain group of artists, which were obtained via chart.getTopArtists endpoint \citep{robinson2020user}.
After removing duplicates from the approximately $1,100,000$ crawled user names, a grand total of almost $400,000$ unique users were harvested.
Due to the substantial number of listening records per user, $10,000$ unique users were randomly chosen.
Through other Last.fm Application Programming Interfaces (APIs), a significant number of unique listening events (LEs) from these users were available.

\subsection{Dataset Description}\label{subsec: data-description}
Play count measures how frequently the observed user engages in music listening \citep{farrahi2014impact}.
For tracks with a high play count, we can assume that the user has likely enjoyed them.
However, tracks that are played rarely or have not been played cannot be easily dismissed as unappealing to users.
Some tracks that users cannot access have a play count of $0$.
Additionally, there are tracks that users do not currently prefer, but it does not imply that they will not prefer them in the long run as their preferences evolve.
The interaction between a user and a track (or artist, album) is reflected in the fact that the user has listened to a particular track, i.e., such a listening event exists.
Referring to the article \citep{robinson2020user}, we also use a simple key consisting of artist and track name tuples to distinguish individual tracks.

The basic statistical information contained in the dataset is shown in Table \ref{tab: basic-dataset-characteristics}.
For this experiment, two Last.fm datasets were utilized, named lastfm-100k and lastfm-200k, according to the approximate size of the datasets.
In the filtered datasets, lastfm-100k contains $501$ users with close to $100,000$ entries for listening events (LEs).
In comparison, the number of users and LEs in lastfm-200k is almost twice as large, with $1,001$ users and more than $200$ thousand LEs.
The number of tracks in the two datasets is $25,279$ and $42,668$, respectively.
Regarding the average actions per user, average actions per track and sparsity, the two datasets are quite similar, with values of $199.3$ vs $200.0$, $4.0$ vs $4.7$, and $99.2$\% vs $99.5$\%, respectively.

\begin{table}[htp]
  \centering
  \caption{Basic dataset characteristics}\label{tab: basic-dataset-characteristics}
  \resizebox{\linewidth}{!}{
  {
  \begin{tabular}{ccccccc}
    \toprule
    \multirow{3}*{Name} & \multicolumn{6}{c}{Quantity} \\
    \cline{2-7}
    & Users  & Tracks & LEs & \makecell[c]{Average actions \\ per user} & \makecell[c]{Average actions \\ per track} & Sparsity \\
    \midrule
    lastfm-100k & 501  & 25,279 & 99,861  & 199.3 & 4.0 & 99.2\%  \\
    lastfm-200k & 1,001 & 42,668 & 200,173 & 200.0 & 4.7 & 99.5\%  \\
    \bottomrule
  \end{tabular}}}
\end{table}

\subsection{Experimental Setup}
The algorithm DWHRec generates the recommendation list by initially calculating the scores of users and candidate tracks according to the customized scoring function, and then ranking the candidate tracks depending on the scores.
To ensure that each user and each track in the testset carries its own vector representation, extra care is required in constructing the dataset.
For each user, we ranked their listening records and spanned the top $200$ records (it was observed that the vast majority of the users' listening records ranked above $200$ had single-digit listening times).
In the experimental section, we stochastically split the dataset into training and test sets, with the training set comprising $90\%$ and the testset containing the remaining $10\%$.
We conducted ten experiments, iterating over all ten permutations, and the final result displayed is the mean of the ten outcomes.
For baselines in our experiment, we evaluated their performance using the RecBole toolkit~\cite{xin2021recbole, xin2022recbole2.0, xu2023towards}.

DWHRec contains six hyper-parameters, the number of iterations $r$ for random walks, the number of steps $k$ for the walks, the representation dimension $s$ of the vector, the word embedding window size $w$, the length of the recommendation list $n$ and the weighting factor $\alpha$ of the scoring function.
In the experiment, the first five hyper-parameters are assigned values using a manually set approach, sequentially set as $r\text{=}5$, $k\text{=}200$, $s\text{=}50$ and $w\text{=}5$.
Regarding the diversity weighting factor \(\alpha\), an adaptive weight is employed, where the diversity weight for the $i$-th item to be recommended is given by $\alpha_i = (1 - \frac{1}{i+1})$.

\subsection{Evaluation Metrics}
Five evaluation metrics, categorized into accuracy and diversity, were applied to compare different algorithms \cite{zanon2022balancing}.
Recall and Hit Ratio, two accuracy metrics unrelated to ordering, measured the fundamental accuracy.
Ranking accuracy was assessed using two metrics: Mean Average Precision (MAP) and Normalized Discounted Cumulative Gain (NDCG).
The Aggregate Diversity (AGGR-DIV) metric was utilized to measure the diversity of recommendations.

The AGGR-DIV metric measures the degree of diversity among items in the recommendation list across all users.
It is computed as the total number of unique genres on tracks in the recommendation list, considering all users.
AGGR-DIV is defined as follows:
\begin{equation}\label{equ: metrics-aggr-div}
\text{AGGR-DIV}(L_u) = \sum_{ta} \left( p(ta) \times D(ta) \right) / \sum_{ta} D(ta), \, ta \in Ta(L_u),
\end{equation}
Where $L_u$ represents the recommendation list for user $u$, and $Ta(L_u)$ represents the set of tags annotated to the tracks in $L_u$.
In Equation \eqref{equ: metrics-aggr-div}, $p(ta)$ is the cumulative discounted probability of $ta$ over $L_u$, and $D(ta)$ represents the number of occurrences of $ta$ in $L_u$.
The notation $d(tr, ta)$ represents the number of occurrences of $ta$ in $tr$, taking a value of $1$ if $ta$ is present in $tr$ and $0$ otherwise.
$D(ta)$ and $d(tr, ta)$ are defined as follows:
\begin{equation*}
  D(ta) = \sum_{tr \in L_u} d(tr, ta), \,
  d(tr, ta) = 
  \begin{cases}
    1, & \mbox{if } tr \text{ has } ta; \\
    0, & \mbox{otherwise}.
  \end{cases}
\end{equation*}
For the track $tr \in L_u$, if $tr$ contains the tag $ta$, the count of $ta$ is denoted as $c(tr, ta)$; otherwise, the count is zero.
Then, the proportion of $ta$ in the tags associated with $tr$ is equal to the frequency of occurrence of $ta$, denoted as $q(tr, ta)$.
The probability of $ta$ in $tr$, expressed as $p(tr, ta)$, is calculated in a discounted form of $q(tr, ta)$,
\begin{equation*}
  p(ta) = \sum_{tr \in L_u} p(tr, ta), \,
  p(tr, ta) = \frac{q(tr, ta)}{\log_2 \left(1 + j \right)}, \,
  q(tr, ta) = c(tr, ta) / \sum_{ta_i \in Ta(tr)} c(tr, ta_i).
\end{equation*}
where $j$ represents that $tr$ is the $j$-th track in $L_u$ that contains $ta$.
By traversing $L_u$ and accumulating $p(tr, ta)$, we obtain the probability $p(ta)$ for $ta$.
A higher AGGR-DIV value indicates better diversity in the recommended results.

\subsection{Baselines}
We compared our approach with several state-of-the-art recommendation algorithms.
Below is a brief description of these algorithms.

\begin{itemize}
  \item \textbf{P}opularity \textbf{B}ased (\textbf{PB}) \cite{rakshit2023popularity}: The popularity-based recommender always recommends to users the $n$ most frequently listened-to tracks by all users in the dataset.
  {
  \item \textbf{B}ayesian \textbf{P}ersonalized \textbf{R}anking (\textbf{BPR}) \cite{rendle2012bpr}:
  It is the most widely used method for recommendation based on matrix factorization with pairwise ranking loss.

  \item \textbf{Neu}ral Collaborative Filtering--\textbf{M}atrix \textbf{F}actorization (\textbf{NeuMF}) \cite{he2017neumf}:
  The author is dedicated to developing neural network-based techniques to tackle the critical issue in recommendation, specifically collaborative filtering based on implicit feedback.
  \item \textbf{D}eep \textbf{M}atrix \textbf{F}actorization (\textbf{DMF}) \cite{xue2017deep}: The algorithm constructs a user-item matrix that incorporates explicit ratings and non-preference implicit feedback.
  It introduces a novel matrix factorization model with a neural network architecture to learn a generic low-dimensional space representation of users and items.
  \item \textbf{Light} \textbf{G}raph \textbf{C}onvolution \textbf{N}etwork (\textbf{LightGCN}) \cite{he2020lightgcn}:
  LightGCN learns user and item embeddings by linearly propagating them on the user-item interaction graph.
  It uses the weighted sum of embeddings learned at all layers as the final embedding.
  \item \textbf{D}iversified \textbf{G}NN-based \textbf{Rec}ommender System (\textbf{DGRec}) \cite{yang2023dgrec}: It is currently the state-of-the-art diversified recommendation algorithm. The authors suggest diversifying GNN-based recommender systems by directly improving the embedding generation procedure.
  }
  \item \textbf{H}ypergraph \textbf{E}mbeddings for \textbf{M}usic \textbf{R}ecommendation (\textbf{HEMR}) \cite{gatta2022music}:
  It is a recommendation algorithm specifically designed for music recommendations, which is the most similar baseline to our method by utilizing hypergraphs.
\end{itemize}

\section{Experimental Results}\label{sec: experimental-results}
Conducting comparative experiments between DWHRec and seven other recommendation algorithms on two datasets, using five metrics to validate recommendation performance, aims to assess the model's effectiveness.
We delve into the analysis of the impact of hyper-parameter variations on the model through sensitivity experiments.
In ablation experiments, we dissect the influence of different types of hyperedges on the model.

\subsection{Overall Comparison}
\label{subsec: overall-comparison}
Seven advanced algorithms, namely PB, BPR, NeuMF, DMF, LightGCN, HEMR and DGRec, are employed as comparative models.
DWHRec is compared with these models using the same experimental datasets, lastfm-100k and lastfm-200k (Table \ref{tab: basic-dataset-characteristics}).
The quality of the recommendation results is evaluated based on metrics, including accuracy indicators such as map, recall, hit ratio and ndcg, as well as diversity metrics represented by aggr-div.

Table \ref{tab: experimental-result-summary} shows the results of the eight recommendation algorithms on two datasets utilizing evaluation metrics such as map, recall, hit ratio, ndcg and aggr-div.
The table is divided into two sections: the upper half showcases results for the lastfm-100k dataset, while the lower half displays results for the lastfm-200k dataset.
Both datasets share identical row and column names.
The row names correspond to the abbreviations for eight comparative models, while the column names represent metrics for a recommended length of $20$.
The values in the table indicate the performance of each model under specific metrics.
Bold entries highlight the best results, entries with an underline (``\_") signify the second-best results, and entries with a wavy underline (``\texttildelow") indicate the third-best results.

\begin{table}[htp]
  \centering
  \caption{Comparison results of our model with different baselines on five evaluation metrics}
  \label{tab: experimental-result-summary}
  \begin{tabular}{lccccc}
    \toprule
    \multirow{2}*{Models} & \multicolumn{5}{c}{Evaluation Metrics} \\
    \cline{2-6}
    & MAP@20 & Recall@20 & Hit Ratio@20 & NDCG@20 & AGGR-DIV@20  \\
    \midrule
    & \multicolumn{5}{c}{lastfm-100k} \\
    \cline{2-6}
    PB       & 0.0002          & 0.0014          & 0.0271          & 0.0077          & \uwave{0.0145}  \\
    BPR      & 0.0002          & 0.0010          & 0.0198          & 0.0067          & 0.0106 \\
    NeuMF    & 0.0014          & 0.0070          & 0.1178          & 0.0425          & 0.0106 \\
    DMF      & \uwave{0.0019}  & \uwave{0.0074}  & \uwave{0.1186}  & \uwave{0.0482}  & 0.0116 \\
    LightGCN & \uline{0.0038}  & \uline{0.0143}  & \uline{0.2052}  & \uline{0.0842}  & 0.0113 \\
    HEMR     & 0.0009          & 0.0048          & 0.0629          & 0.0195          & \uline{0.1181} \\
    DGRec    & 0.0006          & 0.0027          & 0.0437          & 0.0162          & 0.0108 \\
    DWHRec   & \textbf{0.0149} & \textbf{0.0252} & \textbf{0.2559} & \textbf{0.1643} & \textbf{0.1721} \\

    \midrule
    & \multicolumn{5}{c}{lastfm-200k} \\
    \cline{2-6}
    PB       & 0.0002          & 0.0022          & 0.0415          & 0.0114          & \uwave{0.0153} \\
    BPR      & \uwave{0.0036}  & \uwave{0.0109}  & \uwave{0.1276}  & \uwave{0.0583}  & 0.0103 \\
    NeuMF    & \uline{0.0047}  & \uline{0.0204}  & \textbf{0.3095} & \uline{0.1226}  & 0.0113 \\
    DMF      & 0.0005          & 0.0025          & 0.0488          & 0.0173          & 0.0103 \\
    LightGCN & 0.0017          & 0.0060          & 0.0874          & 0.0362          & 0.0113 \\
    HEMR     & 0.0009          & 0.0052          & 0.0672          & 0.0206          & \uline{0.1296} \\
    DGRec    & 0.0015          & 0.0071          & 0.1047          & 0.0389          & 0.0111 \\
    DWHRec   & \textbf{0.0142} & \textbf{0.0220} & \uline{0.2211}  & \textbf{0.1501} & \textbf{0.1842} \\
    \bottomrule
  \end{tabular}
\end{table}

On the lastfm-100k dataset in Table \ref{tab: experimental-result-summary}, it is evident that DWHRec outperformed other algorithms significantly by a considerable margin across the five metrics provided.
DWHRec consistently occupied a leading position under these metrics.
LightGCN also demonstrated strong performance, securing the second position in four other metrics except for aggr-div@20.
HEMR attained the second position in ndcg@20, while PB secured the third position in the same metric.
For NeuMF, its performance closely approached that of DMF.
The remaining algorithms distinctly lagged behind both DMF and NeuMF in performance.

On the lastfm-200k dataset in Table \ref{tab: experimental-result-summary}, changes in trends were observed.
DWHRec exhibited outstanding performance, notably excelling in map@20, ndcg@20 and aggr-div@20 metrics, surpassing other algorithms by a significant margin.
In terms of map@20 and recall@20 metrics, DWHRec, NeuMF and BPR occupied the top tier, securing the first three positions.
The performance of NeuMF and DWHRec on recall@20 was remarkably close, with minimal differences.
LightGCN's performance, however, was not as robust on this dataset.
The distinction between DGRec and LightGCN was relatively small.
In terms of four accuracy metrics, BPR outperformed both LightGCN and DGRec, with the exception of the aggr-div@20 metric.

\begin{figure}[H]
    \centering
    \subfigure[Sub-fig for metric ndcg on lastfm-100k]{
        \label{subfig: experimental-result-ndcg-on-100k}
        \includegraphics[width=0.48\textwidth]{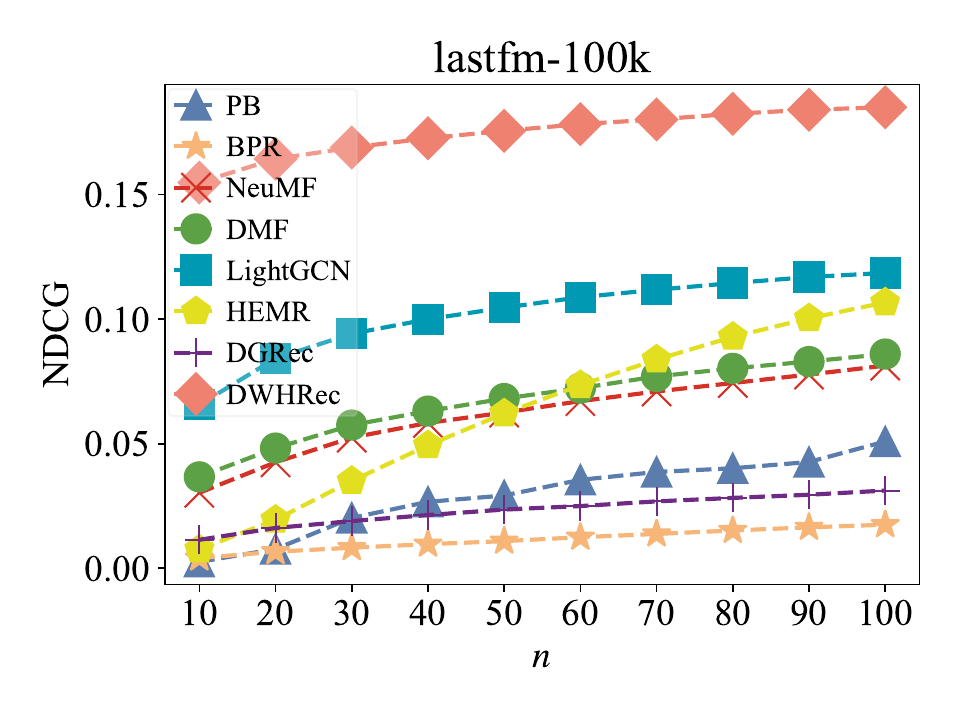}}
    \hspace{0.01cm}
    \subfigure[Sub-fig for metric aggr-div on lastfm-100k]{
        \label{subfig: experimental-result-aggrdiv-on-100k}
        \includegraphics[width=0.48\textwidth]{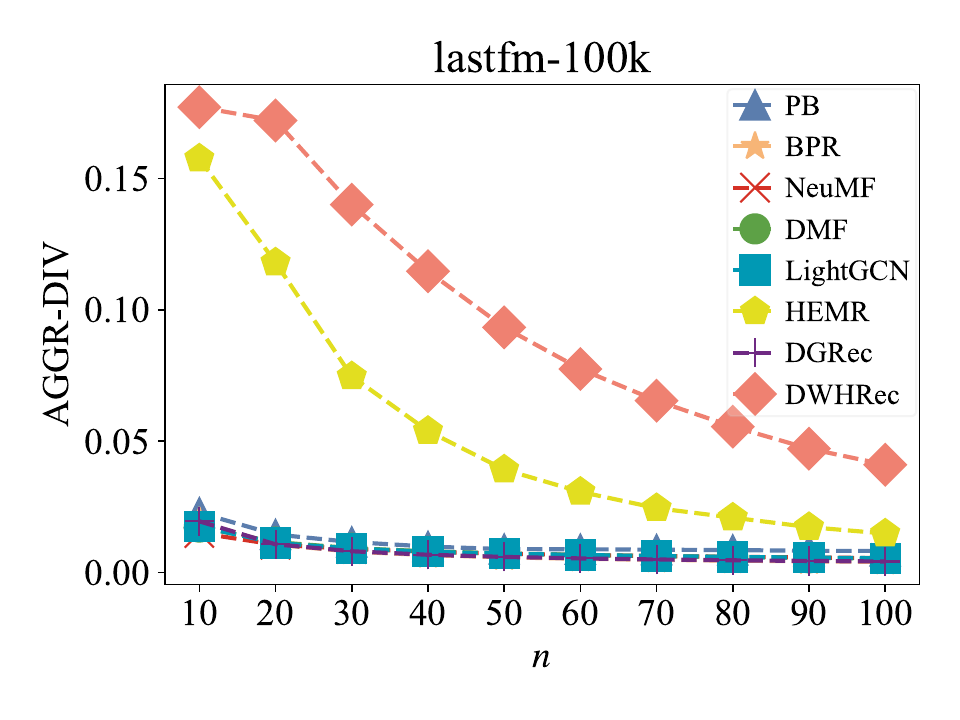}}
    \hspace{0.01cm}
    \subfigure[Sub-fig for metric ndcg on lastfm-200k]{
        \label{subfig: experimental-result-ndcg-on-200k}
        \includegraphics[width=0.48\textwidth]{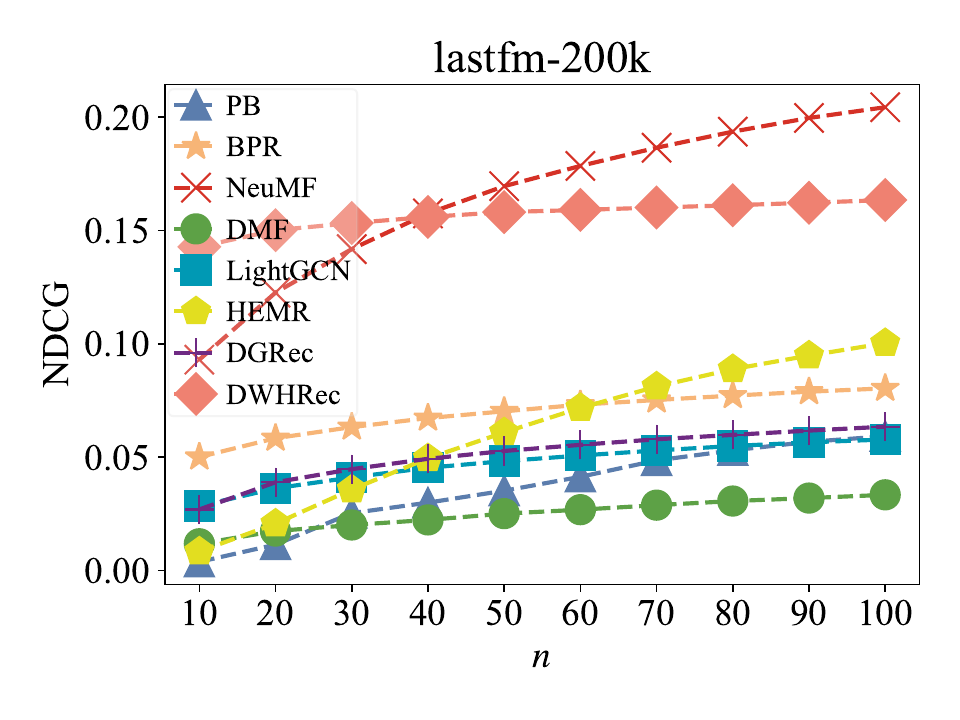}}
    \hspace{0.01cm}
    \subfigure[Sub-fig for metric aggr-div on lastfm-200k]{
        \label{subfig: experimental-result-aggrdiv-on-200k}
        \includegraphics[width=0.48\textwidth]{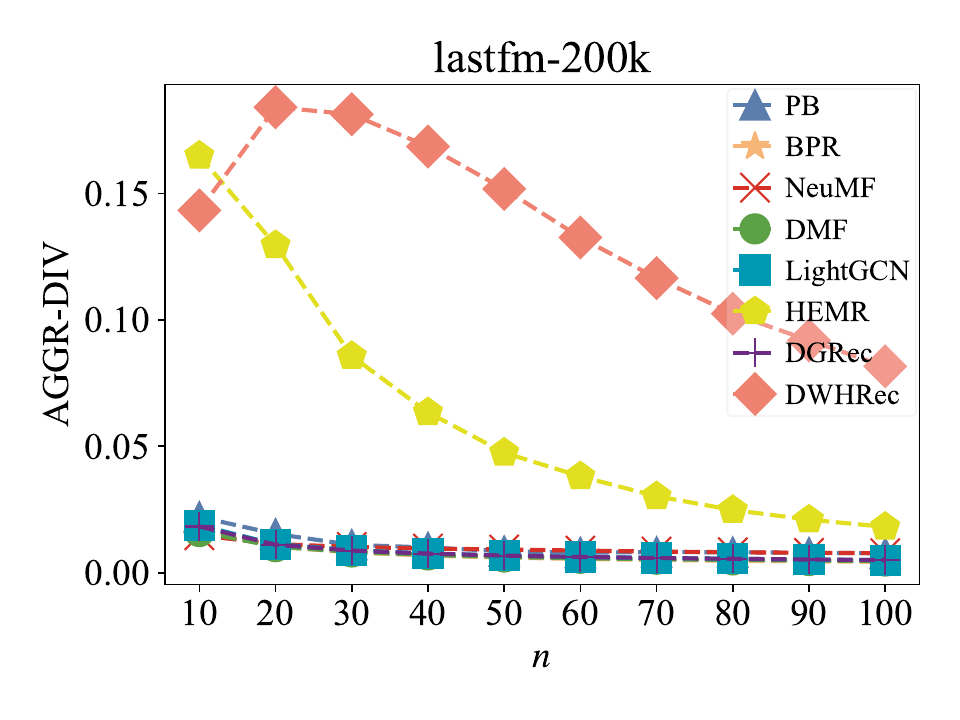}}
  \caption{Summary of comparison results of our model with different baselines on evaluation metrics.}
  \label{fig: experimental-result-summary}
\end{figure}

Figure \ref{fig: experimental-result-summary} proclaims a brief overview of the comparative results, featuring two key metrics: ndcg for accuracy and aggr-div for diversity.
The horizontal axis of the figure depicts the length of the recommendation list ($n$), spanning ten data points from $10$ to $100$.
To enhance visibility, all lines are uniformly represented with dashed lines (``\text{-}\text{-}"), each accompanied by distinct colors and point shapes.
Notably, the color of each line matches the color of the points along it.
Blue-gray triangles denote PB (``\textcolor[HTML]{5B7DAD}{\trianglepafill}").
Light pink stars represent BPR (``\textcolor[HTML]{F7B577}{\starletfill}").
NeuMF is marked with red cross symbols (``\textcolor[HTML]{D53126}{$\times$}").
DMF is identified by olive circular points (``\textcolor[HTML]{5DA146}{\circletfill}").
Light blue colored squares fill the LightGCN (``\textcolor[HTML]{0099B4}{\squadfill}").
The golden pentagons make up the HEMR (``\textcolor[HTML]{E2DE20}{\pentagofill}").
Purple plus signs characterize the DGRec (``\textcolor[HTML]{6F2A83}{\linevh}").
DWHRec is adorned with orange diamonds (``\textcolor[HTML]{EF8171}{\rhombusfill}").

Figure \ref{fig: experimental-result-summary} is composed of four parts labeled a, b, c and d.
Figures \ref{subfig: experimental-result-ndcg-on-100k} and \ref{subfig: experimental-result-aggrdiv-on-100k} depict the ndcg and aggr-div results for the lastfm-100k dataset, while Figures \ref{subfig: experimental-result-ndcg-on-200k} and \ref{subfig: experimental-result-aggrdiv-on-200k} portray the corresponding results for the lastfm-200k dataset.
All four figures share a common horizontal axis representing the values of $n$. 
The vertical axis is organized into two groups: the ndcg metric for Figures \ref{subfig: experimental-result-ndcg-on-100k} and \ref{subfig: experimental-result-ndcg-on-200k}, and the aggr-div metric for Figures \ref{subfig: experimental-result-aggrdiv-on-100k} and \ref{subfig: experimental-result-aggrdiv-on-200k}.

Across the $n$ range from $10$ to $100$, the ndcg results for all algorithms increased as $n$ grew (Figures \ref{subfig: experimental-result-ndcg-on-100k} and \ref{subfig: experimental-result-ndcg-on-200k}).
However, the growth rates of ndcg values varied among the algorithms.
In the lastfm-100k dataset (Figure \ref{subfig: experimental-result-ndcg-on-100k}), HEMR exhibited the fastest growth rate.
DWHRec, DMF, NeuMF, LightGCN and PB accelerated at comparable rates, forming the second tier.
The remaining algorithms demonstrated slower growth rates.
Whereas in the lastfm-200k dataset (Figure \ref{subfig: experimental-result-ndcg-on-200k}), NeuMF displayed the sharpest growth rate, with HEMR closely following its pace and showing a commendable growth spurt.
Over the range of $n$ from $10$ to $100$, the aggr-div results for all algorithms decreased with the increase in $n$ (Figures \ref{subfig: experimental-result-aggrdiv-on-100k} and \ref{subfig: experimental-result-aggrdiv-on-200k}).
As $n$ increases, the aggr-div values for all algorithms remain closely aligned, except for HEMR and DWHRec.

Throughout the entire experiment, DWHRec consistently achieved favorable results in evaluation metrics such as map@20, recall@20, ndcg@20 and aggr-div@20, yielding significant differences from the other seven algorithms (Table \ref{tab: experimental-result-summary} and Figure \ref{fig: experimental-result-summary}).
Notably, DWHRec delivered commendable aggr-div outcomes, which is an encouraging result.

\subsection{Hyper-parameter Sensitivity}
The DWHRec model includes multiple hyper-parameters, roughly categorized into two main types.
The hyper-parameter $r$ represents the number of iterations for random walks, and $k$ denotes the number of steps a vertex can take in a single iteration.
Hyper-parameters $r$ and $k$ control the random walks state of vertices in the hypergraph.
Hyper-parameter $w$ is the size of the sliding window, and $s$ is the dimension of the representation vector in the word embedding process.
Hyper-parameters $s$ and $w$ intervene in the word embedding process, thereby influencing the generation of vertex vectors.
Through sensitivity experiments on these hyper-parameters, we aim to explore how their variations impact the final recommendation performance.

\begin{figure}[H]
    \centering
    \subfigure[Sub-fig for metric recall]{
        \label{subfig: experimental-result-hyperparameter-k-recall}
        \includegraphics[width=0.48\textwidth]{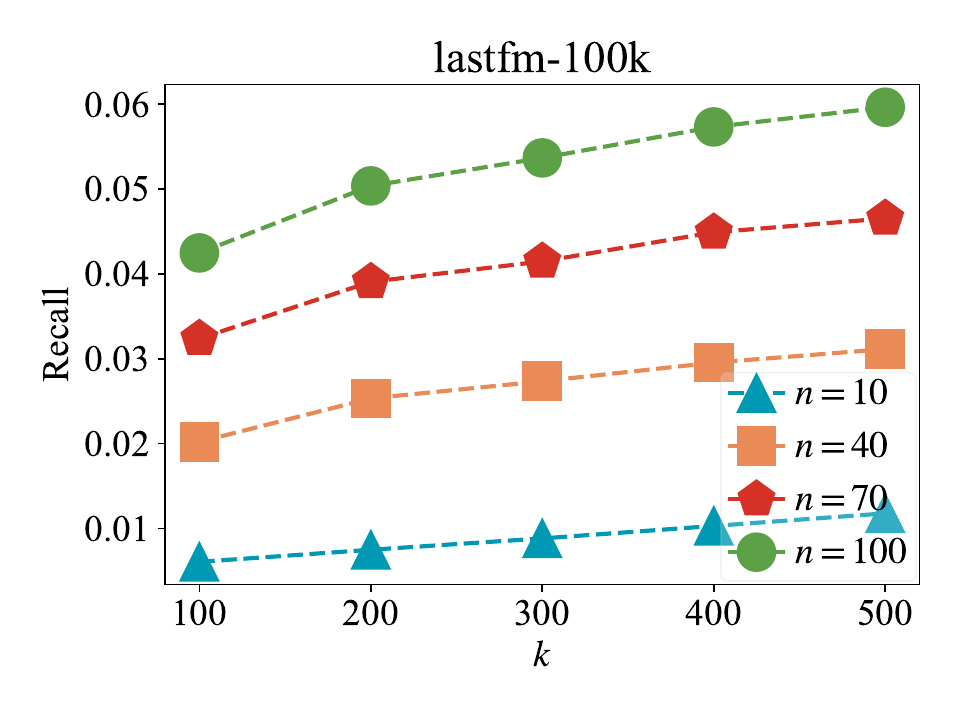}}
    \hspace{0.01cm}
    \subfigure[Sub-fig for metric aggr-div]{
        \label{subfig: experimental-result-hyperparameter-k-aggr-div}
        \includegraphics[width=0.48\textwidth]{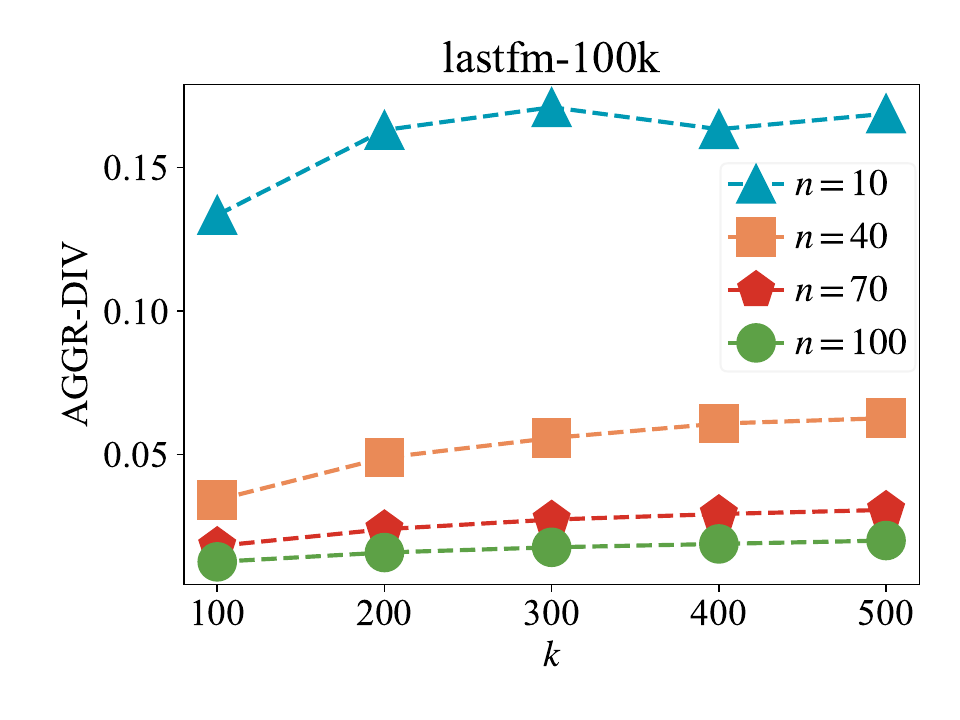}}
  \caption{Sensitivity of hyper-parameter $k$. $k$ denotes the number of steps a vertex taken in each iteration of random walk.}
  \label{fig: experimental-result-hyperparameter-k}
\end{figure}

\begin{figure}[H]
    \centering
    \subfigure[Sub-fig for metric recall]{
        \label{subfig: experimental-result-hyperparameter-r-recall}
        \includegraphics[width=0.48\textwidth]{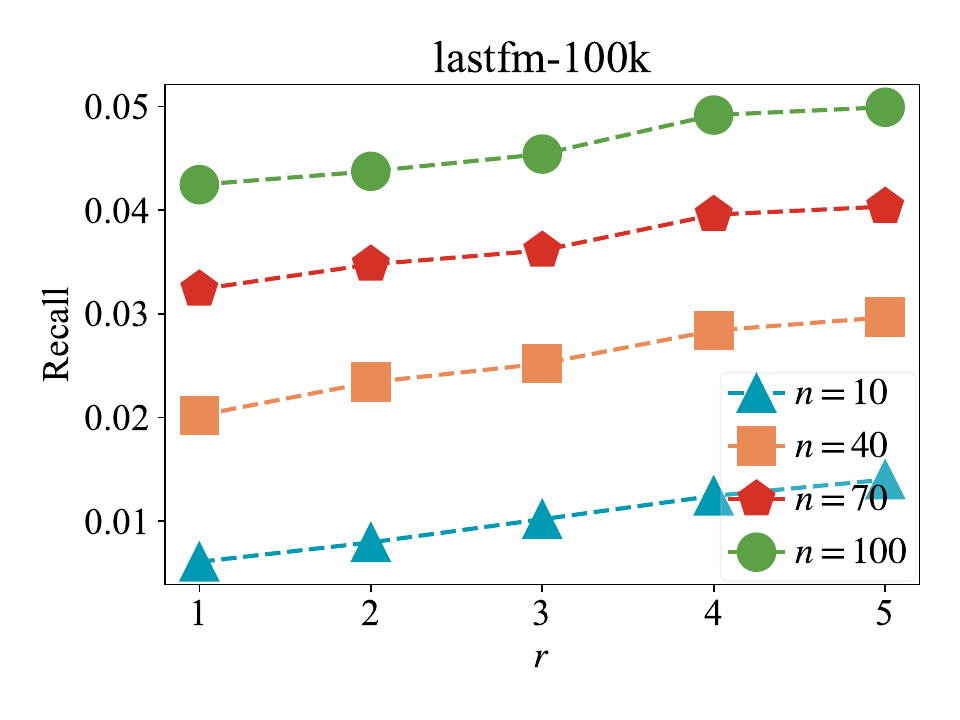}}
    \hspace{0.01cm}
    \subfigure[Sub-fig for metric aggr-div]{
        \label{subfig: experimental-result-hyperparameter-r-aggr-div}
        \includegraphics[width=0.48\textwidth]{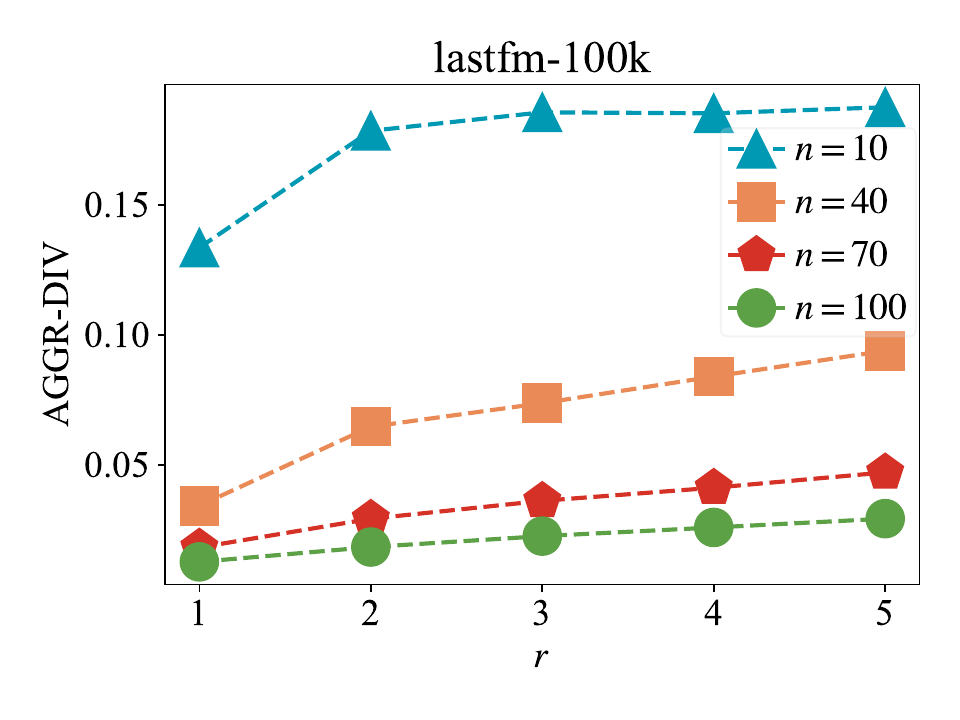}}
  \caption{Sensitivity of hyper-parameter $r$. $r$ controls the number of iterations for random walk on vertices.}
  \label{fig: experimental-result-hyperparameter-r}
\end{figure}

\begin{figure}[H]
    \centering
    \subfigure[Sub-fig for metric recall]{
        \label{subfig: experimental-result-hyperparameter-s-recall}
        \includegraphics[width=0.48\textwidth]{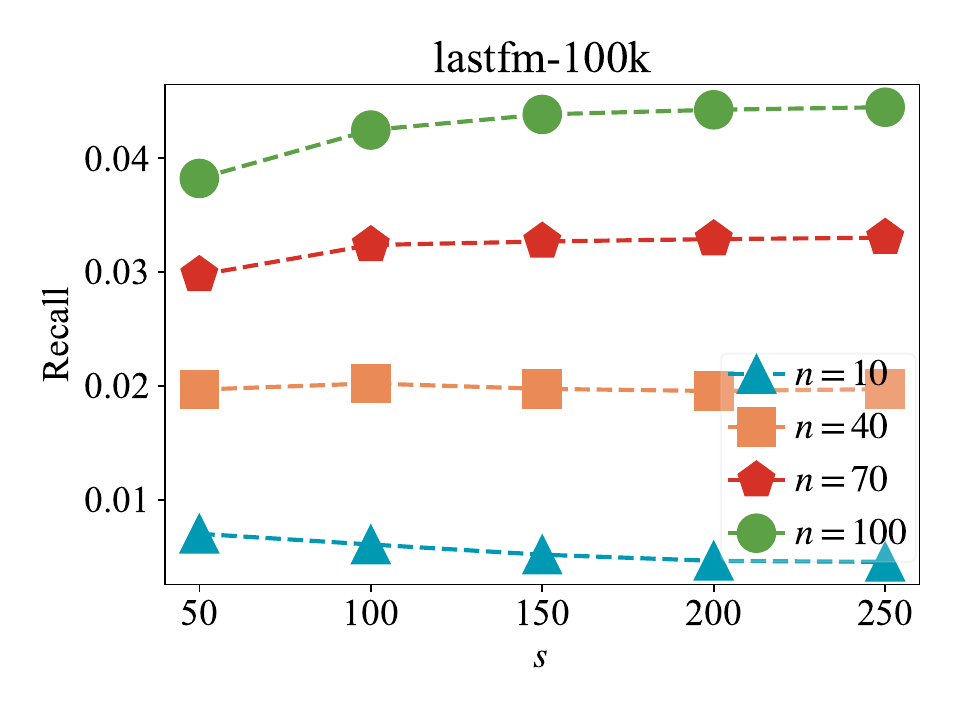}}
    \hspace{0.01cm}
    \subfigure[Sub-fig for metric aggr-div]{
        \label{subfig: experimental-result-hyperparameter-s-aggr-div}
        \includegraphics[width=0.48\textwidth]{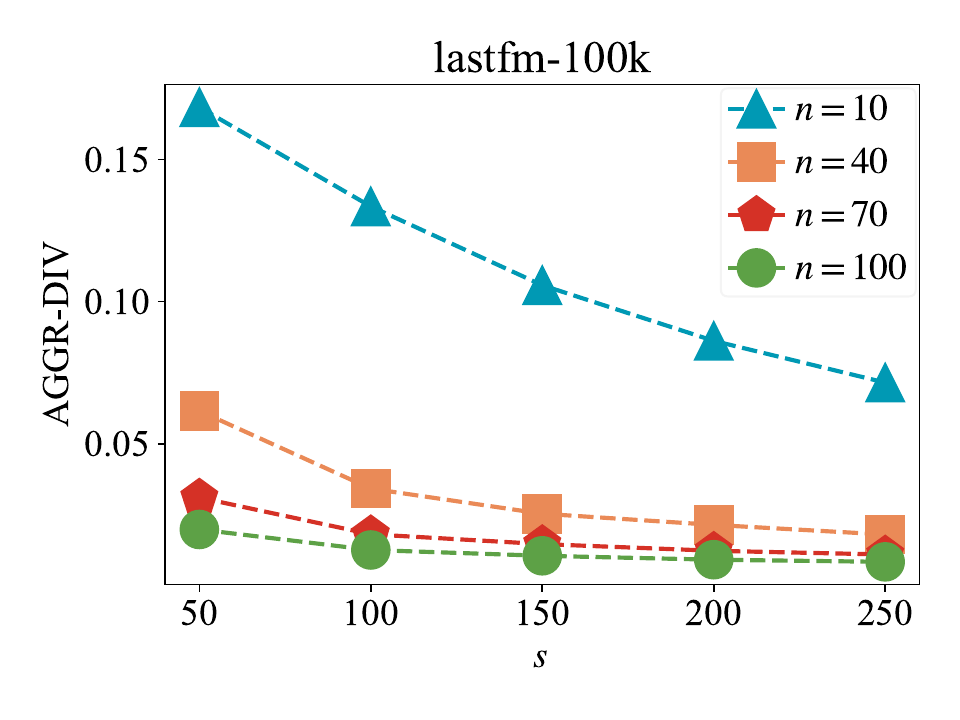}}
  \caption{Sensitivity of hyper-parameter $s$. $s$ adjusts the dimension of the vector representation for vertices.}
  \label{fig: experimental-result-hyperparameter-s}
\end{figure}

\begin{figure}[H]
    \centering
    \subfigure[Sub-fig for metric recall]{
        \label{subfig: experimental-result-hyperparameter-w-recall}
        \includegraphics[width=0.48\textwidth]{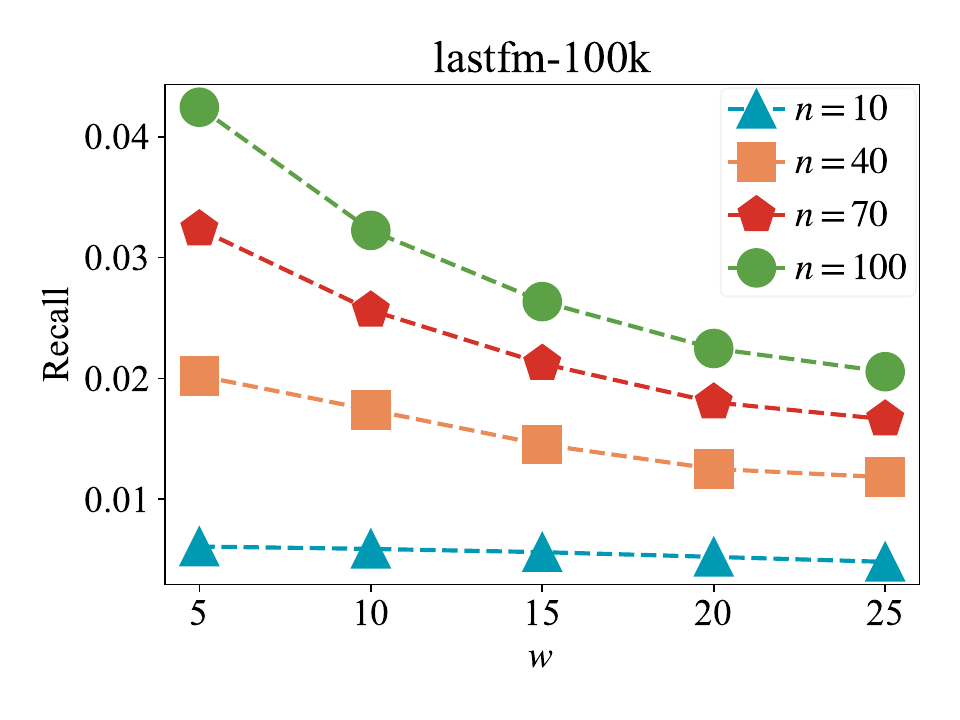}}
    \hspace{0.01cm}
    \subfigure[Sub-fig for metric aggr-div]{
        \label{subfig: experimental-result-hyperparameter-w-aggr-div}
        \includegraphics[width=0.48\textwidth]{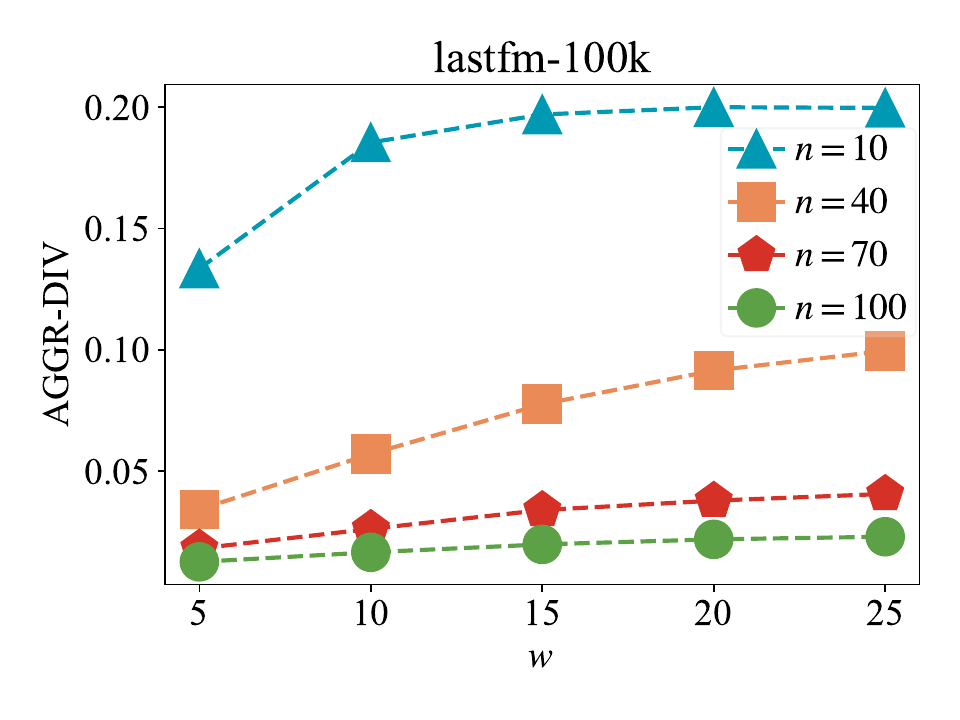}}
  \caption{Sensitivity of hyper-parameter $w$. $w$ adjusts the dimension of the vector representation for vertices.}
  \label{fig: experimental-result-hyperparameter-w}
\end{figure}

For the sensitivity experiments on hyper-parameters, some preparatory work was undertaken.
Specifically, the lastfm-100k dataset was selected as the experimental subject.
Four recommendation lengths were chosen, namely $n\text{=}10$, $40$, $70$ and $100$.
Here, $n\text{=}100$ represents a scenario with a sufficiently long recommendation length, $n\text{=}10$ signifies an extremely short length, while $n\text{=}30$ and $70$ represent typical scenarios.
During the testing of a specific hyper-parameter, other hyper-parameters were kept constant.
Two metrics, recall and aggr-div, were used to characterize the performance of the recommendations.
The experimental results are shown in Figures \ref{fig: experimental-result-hyperparameter-k}--\ref{fig: experimental-result-hyperparameter-w}.

In practice, when conducting sensitivity experiments for hyper-parameters, attention to certain details is crucial.
Both $k$ and $r$ are hyper-parameters that affect the random walks process (Algorithm \ref{algorithm: random-walks-generation}).
To investigate the impact of $k$, $r$ is kept constant at a value of $1$.
Likewise, when testing $r$, $k$ is fixed at $100$.
The other two hyper-parameters, $s$ and $w$, are set to values of $100$ and $5$, respectively.
Both $s$ and $w$ are hyper-parameters that influence the word embedding process.
To explore the impact of $s$ on performance, $w$ is set to a common value of $5$.
Conversely, when investigating the influence of $w$, $s$ is set to a constant value of $100$.
As for the remaining two hyper-parameters, $r$ and $k$, they are kept as small as possible and constant, with specific values of $1$ and $100$, respectively.
The aforementioned configurations are designed to objectively and accurately assess the impact of hyper-parameters on model performance.

The experimental results illustrate that the choice of $k$ has a noticeable impact on the model's performance (Figure \ref{fig: experimental-result-hyperparameter-k}), whether in terms of recall (Figure \ref{subfig: experimental-result-hyperparameter-k-recall}) or aggr-div (Figure \ref{subfig: experimental-result-hyperparameter-k-aggr-div}).
Under the recall metric (Figure \ref{subfig: experimental-result-hyperparameter-k-recall}), a longitudinal view reveals that for the same set of $k$ values, a larger $n$ value corresponds to a larger recall value.
Observing along the horizontal axis shows that, for the four selected values of $n$, as $k$ increases, the recall values also increase.
However, different values of $n$ result in varying degrees of improvement in recall.
Given a change of $100$ in $k$ as one unit, when $n$ takes the minimum, intermediate and maximum values, the average increase in recall metric values is $18.07$\%, $10.68$\% and $9.0$\%, respectively, for each unit increase in $k$.
For the aggr-div metric (Figure \ref{subfig: experimental-result-hyperparameter-k-aggr-div}), a similar trend is observed on the horizontal axis as in the recall metric.
When $n$ is set to $10$, $40$, $70$ and $100$, the aggr-div values increase by $6.47$\%, $12.49$\% and $15.90$\%, respectively, as $k$ increases by one unit.
In contrast to the recall metric, in the aggr-div indicator, the average growth rate tends to increase with larger $n$ values.
Additionally, in the vertical direction, the pattern of aggr-div values is opposite to that of the recall metric.
For the same $k$ value, smaller $n$ values result in larger aggr-div values.

The impact of the hyper-parameter $r$ on the model's recommendation performance is very similar to that of $k$ (Figure \ref{fig: experimental-result-hyperparameter-r}).
The effects of changing $n$ within the same set of $r$ values and the influence of different $r$ values on the metrics are consistent with the conclusions under $k$ values.
The main difference introduced by the two hyper-parameters is reflected in the rate of change of the metric values.
Compared to $k$, $r$ yields a greater increase in recall values (Figure \ref{subfig: experimental-result-hyperparameter-r-recall}) when $n$ is at its minimum, with a growth rate of $23.42$\%.
However, when $n$ takes intermediate and maximum values, the growth rates of recall values are lower than the performance under $k$, at $7.93$\% and $4.16$\%, respectively.
Alternatively, concerning the aggr-div metric (Figure \ref{subfig: experimental-result-hyperparameter-r-aggr-div}), the impact of $r$ values is more profound than that of $k$ values.
The growth rates are $9.68$\%, $30.13$\% and $23.91$\% for the three different scenarios of $n$ values, respectively.

The impact of hyper-parameter $s$ on the model's performance is quite intricate (Figure \ref{fig: experimental-result-hyperparameter-s}).
Let $s$ vary by $50$ as a unit.
When $s$ increases by one unit, it brings an average negative growth of $10.11$\% to the recall metric (Figure \ref{subfig: experimental-result-hyperparameter-s-recall}) for a recommended list length of $10$.
However, when $n$ is $100$, it brings an average positive growth rate of $3.95$\%.
In the scenario where $n$ takes the middle value, the improvement of $s$ has a minimal impact on the recall value, with an average growth change of only $1.39$\%.
Across different values of $n$, the changing trend of $s$ on the aggr-div metric remains consistent (Figure \ref{subfig: experimental-result-hyperparameter-s-aggr-div}).
Among them, the impact of $s$ is very close when $n$ takes the minimum and maximum values.
With each unit increase in $s$, aggr-div values decrease by $19.25$\% and $18.38$\%, respectively.
The effect of $s$ is most pronounced when $n$ takes the middle value, causing a decrease of $23.55$\% in aggr-div for each additional unit.

The impact of $w$ on the model is reflected differently in recall and aggr-div (Figure \ref{fig: experimental-result-hyperparameter-w}).
Overall, with the increase of $w$, the recall metric experiences a negative impact (Figure \ref{subfig: experimental-result-hyperparameter-w-recall}), while aggr-div receives a positive influence (Figure \ref{subfig: experimental-result-hyperparameter-w-aggr-div}).
The impact of $w$ on the recall metric strengthens with the increase of the recommended list length.
Assuming $w$ varies in units of $50$, when $n$ is equal to $100$, an increase of one unit in $w$ leads to an average decrease of $16.4$\%.
On the aggr-div metric, an increase in $w$ results in more diversified recommendations.
The average growth rate of aggr-div, induced by the variation in $w$, exhibits an inverted U-shaped trend with the increase in $n$.
When $n$ takes values within the middle range, the average growth rate reaches $27.75$\%.

Throughout the entire sensitivity experiment, we aim to uncover certain patterns:
(1) In conclusion, all four hyper-parameters can influence the model's performance, impacting either the recall or aggr-div metrics, or both.
Specifically, when $r$, $k$, $s$ or $w$ are small, increasing their values leads to noticeable changes.
However, as the hyper-parameter values continue to increase and reach a certain threshold, the model's performance tends to plateau, and in some cases, it may even produce the opposite effect.
(2) To summarize, increasing the value of $k$ and $r$ can enhance the predictive performance of the model within a certain range.
A larger improvement is observed in the recall metric when $n$ is smaller, while a greater enhancement is seen in the aggr-div metric when $n$ is larger.
(3) The impact of $s$ on the recall metric is complex and overall quite subtle.
However, increasing the value of $s$ has a noticeable negative impact on the aggr-div metric.
Conversely, $w$ can balance the recall and aggr-div metrics.
Increasing $w$ can lead to a decrease in recommendation accuracy while simultaneously enhancing the diversity of recommendations.
(4) Under the same set of hyper-parameter values, the larger $n$ is, the larger the recall metric value is.
This is because guessing the user's preferred items in a shorter list is more challenging.
As the recommended list lengthens, it undoubtedly increases the likelihood of the recommended items within the user's preferences.
(5) When each parameter surpasses its respective threshold, the model exhibits strong stability.

\subsection{Ablation Study}
DWHRec model is divided into three key steps: hypergraph construction, random walks and word vector embedding.
Among them, random walks and word vector embedding are designed to obtain computationally convenient user and track vectors.
The core of the model lies in its hypergraph structure.
Investigating the roles of various types of hyperedges can help clarify their relationships and contributions to the model.

In Section \ref{subsec: hypergraph-composition}, four types of hyperedges are introduced in the DWHRec algorithm, denoted as $e^{(1)}$, $e^{(2)}$, $e^{(3)}$ and $e^{(4)}$.
The primary objective of DWHRec is to recommend a more diverse set of tracks to users.
Users and tracks are the two most important atoms.
While the listening events of users to tracks, represented by $e^{(1)}$, serving as a crucial bridge connecting the two, is indispensable.
Therefore, to examine the roles of various types of hyperedges in the model, an ablation study was conducted on the lastfm-100k dataset by alternatively eliminating one of $e^{(2)}$, $e^{(3)}$ and $e^{(4)}$.

\begin{figure}[H]
    \centering
    \subfigure[Sub-fig for metric recall]{
        \label{subfig: ablation-experimental-result-recall}
        \includegraphics[width=0.48\textwidth]{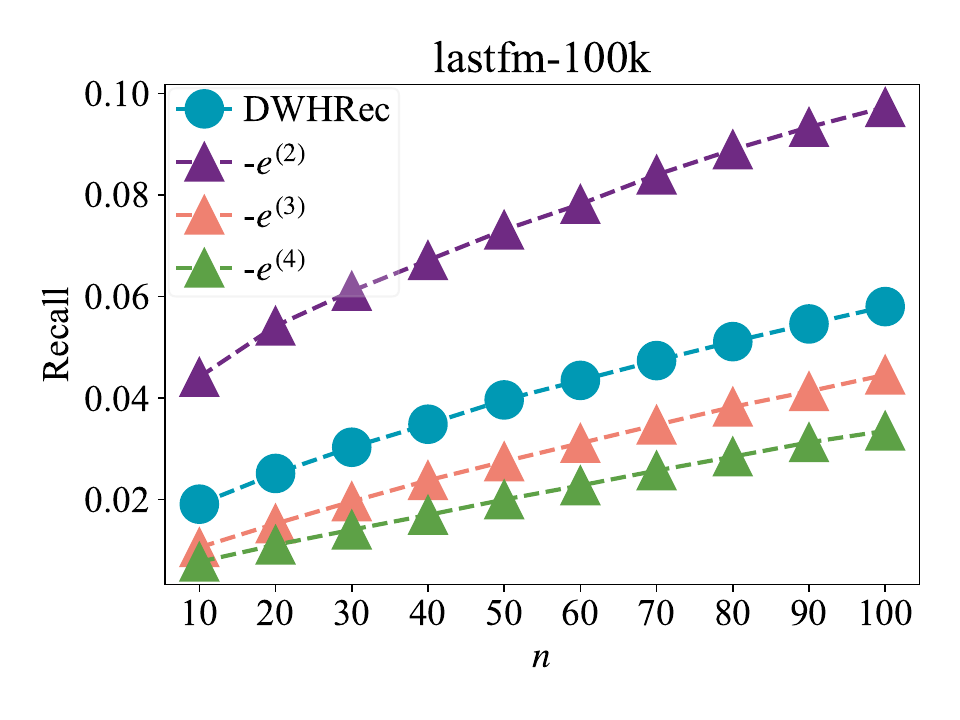}}
    \hspace{0.01cm}
    \subfigure[Sub-fig for metric aggr-div]{
        \label{subfig: ablation-experimental-result-aggr-div}
        \includegraphics[width=0.48\textwidth]{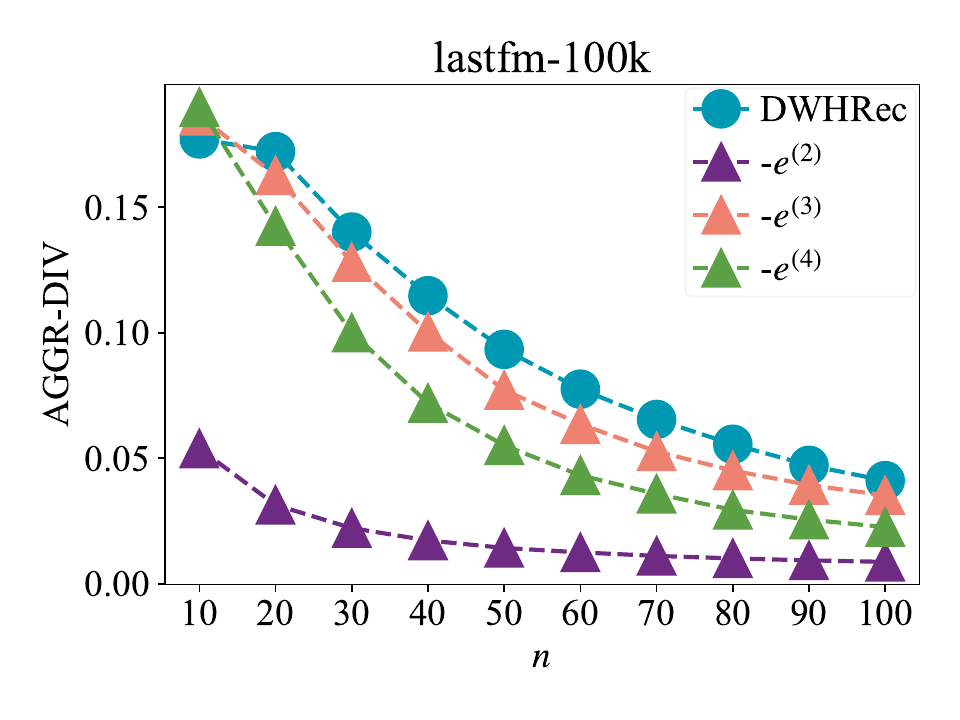}}
  \caption{Complete results of the ablation experiments for various values of $n$}
  \label{fig: experimental-result-ablation}
\end{figure}

The concise experimental results are presented in Table \ref{tab: result-of-ablation-study}, while Figure \ref{fig: experimental-result-ablation} offers a performance comparison of the model based on the recall and aggr-div metrics for all $n$ values.
Table \ref{tab: result-of-ablation-study} is a subset of Figure \ref{fig: experimental-result-ablation}, specifically representing the case when $n=50$.
The prefix symbol ``-" in both Table \ref{tab: result-of-ablation-study} and Figure \ref{fig: experimental-result-ablation} signifies ``remove" or ``subtract", meaning the action of elimination or deletion.

\begin{table}[htp]
    \centering
    \caption{Results of the ablation experiments on the lastfm-100k dataset. We show DWHRec's performance when removing each of the modules.}
    \setlength{\tabcolsep}{10mm}{
    \begin{tabular}{ccc}
         \toprule
         Method & Recall@50 & AGGR-DIV@50 \\
         \midrule
         DWHRec      & 0.0396 & 0.0934 \\
         -$e^{(2)}$  & 0.0731 & 0.0143 \\
         -$e^{(3)}$  & 0.0275 & 0.0772 \\
         -$e^{(4)}$  & 0.0200 & 0.0551 \\
         \bottomrule
    \end{tabular}}
    \label{tab: result-of-ablation-study}
\end{table}

Observing Table \ref{tab: result-of-ablation-study} and Figure \ref{fig: experimental-result-ablation}, we made several findings:
(1) Regarding the recall@50 metric, removing $e^{(3)}$ and $e^{(4)}$ led to a decrease in performance, but removing $e^ {(2)}$ resulted in a significant improvement.
The experiments indicated that compared to the DWHRec model with all types of hyperedges as a baseline, removing $e^{(2)}$, $e^{(3)}$ and $e^{(4)}$ resulted in an increase of $84.60$\%, $-30.56$\% and $-49.49$\%.
In which case, the model's performance, after removing $e^{(3)}$ and $e^{(4)}$ respectively, was only $69.44$\% and $50.51$\% of the full model's performance.
(2) In the case of the aggr-div@50 metric, removing any type of hyperedge seemed to have a clear impact.
Experiments revealed that after removing $e^{(2)}$, $e^{(3)}$ and $e^{(4)}$, the model provided $15.31$\%, $82.66$\% and $58.99$\% of the complete model's performance, with gaps of $84.69$\%, $17.34$\% and $41.01$\%, respectively.
(3) Removing any hyperedge resulted in an apparent drop in aggr-div values, while interventions on recall remained relatively complex.

The ablation experiments analyzed the impact of different types of hyperedges on the model.
The experiments indicated that both the album hyperedge $e^{(3)}$ and the artist hyperedge $e^{(4)}$ could improve the model's performance, whether in terms of recommendation accuracy or diversity.
Comparatively, the impact of $e^{(4)}$ was slightly higher than that of $e^{(3)}$.
Artists might not have released albums but would certainly create tracks.
Therefore, the information captured by $e^{(4)}$ was more comprehensive, while $e^{(3)}$ might have missed some important information.
The tagging hyperedge $e^{(2)}$ tended to interfere with the model's ability to make accurate predictions.
Since tags are directly related to the category of tracks, they could lead to a significant increase in diversity.
Overall, all four types of hyperedges contributed to achieving a balance between recommendation accuracy and diversity.

\section{Conclusions}\label{sec: conclusions}

We developed a novel algorithm, DWHRec, for diversified music recommendations using a weighted hypergraph embedding technique.
This algorithm operates in two stages: constructing a hypergraph from user listening history, tracks, albums, artists and tags; and then recommending tracks by exploring connections through a random walks hypergraph embedding.
We validated DWHRec's effectiveness by comparing it with seven advanced algorithms using a dataset from Last.fm, demonstrating significant improvements across multiple evaluation metrics.
Additionally, we assessed the sensitivity to hyper-parameters and the impact of different types of hyperedges on accuracy and diversity.
Our findings suggest that DWHRec is not only effective for music recommendation but also adaptable to other fields with similar data structures.
The hypergraph approach, particularly the use of various hyperedges, plays a crucial role in enhancing recommendation accuracy and ensuring diversity.

\section*{Acknowledgement}
This work was supported by the Fundamental Research Funds for the Central Universities (Grant No. 2023ZY-SX019).

\bibliography{ref}
\bibliographystyle{model5-names}

\end{document}